\documentclass{aastex63}
\pdfoutput=1

\usepackage{graphicx,epstopdf}
\usepackage{savesym}
\savesymbol{tablenum}
\usepackage{siunitx}
\usepackage{amsmath}
\usepackage{enumerate}
\restoresymbol{SIX}{tablenum}
\usepackage[bookmarks=false,         
     pdfnewwindow=true,      
     colorlinks=true,    
     linkcolor=xlinkcolor,     
     citecolor=xlinkcolor,     
     filecolor=xlinkcolor,  
     urlcolor=xlinkcolor,      
     final=true,
 ]{hyperref}
\hypersetup{
    colorlinks=true,
    linkcolor=blue,
    citecolor=black,
    filecolor=black,
    urlcolor=blue,
}
\received{June 1, 2019}
\revised{January 10, 2019}
\accepted{\today}
\submitjournal{ApJ}

\shorttitle{Low energy electron and positron cosmic ray spectra}
\shortauthors{Mechbal et al}

\begin{document}

\title{Measurement of low-energy cosmic-ray electron and positron spectra at 1 AU with the AESOP-Lite spectrometer}
\author[0000-0002-8579-3964]{Sarah Mechbal}
\affiliation{Santa Cruz Institute for Particle Physics, \\
University of California Santa Cruz, \\
Santa Cruz, CA 95064, USA}
\correspondingauthor{Sarah Mechbal}
\email{smechbal@ucsc.edu}

\author[0000-0003-4865-6968]{Pierre-Simon Mangeard}
\affiliation{Bartol Research Institute, \\
University of Delaware, \\
Newark, DE 19716, USA}

\author{John M. Clem}
\affiliation{Bartol Research Institute, \\
University of Delaware, \\
Newark, DE 19716, USA}

\author[0000-0001-7929-810X]{Paul A. Evenson}
\affiliation{Bartol Research Institute, \\
University of Delaware, \\
Newark, DE 19716, USA}

\author[0000-0002-7850-3711]{Robert P. Johnson}
\affiliation{Santa Cruz Institute for Particle Physics, \\
University of California Santa Cruz, \\
Santa Cruz, CA 95064, USA}

\author{Brian Lucas}
\affiliation{Bartol Research Institute, \\
University of Delaware, \\
Newark, DE 19716, USA}

\author{James Roth}
\affiliation{Bartol Research Institute, \\
University of Delaware, \\
Newark, DE 19716, USA}



\begin{abstract}

We report on a new measurement of the cosmic ray (CR) electron and positron spectra in the energy range of 20 MeV -- 1 GeV. The data were taken during the first flight of the balloon-borne spectrometer AESOP-Lite (Anti Electron Sub Orbital Payload), which was flown from Esrange, Sweden, to Ellesmere Island, Canada, in May 2018. The instrument accumulated over 130 hours of exposure at an average altitude of 3~g.cm$^{-2}$ of residual atmosphere. The experiment uses a gas Cherenkov detector and a magnetic spectrometer, consisting of a permanent dipole magnet and silicon strip detectors (SSDs), to identify particle type and measure the rigidity. Electrons and positrons were detected against a background of protons and atmospheric secondary particles. The primary cosmic ray spectra of electrons and positrons, as well as the re-entrant albedo fluxes, were extracted between 20 MeV -- 1 GeV during a positive solar magnetic polarity epoch. The positron fraction below 100 MeV appears flat, suggesting diffusion dominated solar modulation at low rigidity. The all-electron spectrum is presented and compared with models from a heliospheric numerical transport code. 

\end{abstract}

\section{Introduction} \label{sec:intro}

Positrons and electrons constitute only 1\% of galactic cosmic rays (GCRs). Although their contribution is small compared to that of cosmic ray nuclei, the leptonic component is nevertheless important in understanding the origin and propagation of cosmic rays in the Galaxy and the Heliosphere. Electrons and positrons undergo energy loss processes that nuclei do not, such as synchrotron radiation in magnetic fields, bremsstrahlung energy loss with interstellar gases, and inverse Compton scattering with ambient photons. Their measurement thus provides additional information about their transport and origin that cannot be known from the hadronic component. \newline \indent
Both electron and positron cosmic rays can be produced in the interaction between cosmic ray nuclei and the interstellar matter: that contribution is then referred to as \textit{GCR secondary}. The \textit{GCR primary} contribution of cosmic ray electrons has been known since their discovery in the 1960s~\citep{earl_cloud-chamber_1961,meyer_electrons_1961}, their origin most likely pointing to an acceleration in astrophysical shocks in supernova remnants~\citep{abdo_observation_2010,ackermann_detection_2013}. Before the paradigm-shifting measurements made by PAMELA, AMS-02 and Fermi-LAT~\citep{adriani_anomalous_2009,aguilar_first_2013,the_fermi_lat_collaboration_measurement_2012}, it was long believed that all positrons were of purely secondary origin ~\citep{moskalenko_production_1998}. However, the discovery of an excess at energies above 10 GeV seriously challenged that assumption, pointing to possible new sources, such as pulsars or dark matter particles~\citep{hooper_pulsars_2009,cholis_case_2009}. 
To the knowledge of the authors, the positron fraction has only been measured below 100 MeV  during a negative heliospheric polarity period in the 1960's~\citep{beuermann_cosmic-ray_1969}. 
\newline \indent
In order to interpret any simultaneous measurements of electron and positron cosmic rays done at 1 AU at a given solar epoch, we study these results through the lens of the dynamical heliosphere structure. As cosmic rays enter the boundary of the heliosphere, at $\sim$ 120~AU~\citep{stone_voyager_2013,stone_cosmic_2019}, they encounter the turbulent outward flowing solar wind embedded with the Sun’s heliospheric
magnetic field (HMF). As particles propagate along magnetic field lines, they undergo some major modulation mechanisms: convection and adiabatic energy loss in the expanding solar wind, particle diffusion, and drifts due to the HMF~\citep{jokipii_effects_1977}. The lower the rigidity of the particle, the more susceptible to the mechanisms of solar modulation it will be (the effects of solar modulation are negligible above a few tens of GV). Moreover, because of the well-established 11-year solar cycle, the intensity of CRs on Earth changes with solar activity: when a solar cycle reaches its maximum activity, so do the tilt angle of the heliospheric current sheet (HCS) and the magnitude of the HMF, thus maximizing the suppression of cosmic rays reaching Earth. The inverse scenario occurs during solar minimum activity. \newline \indent
On top of the time-varying nature of the solar modulation, GCRs also encounter gradients and curvatures in the large scale HMF, and the effect of the HCS, causing them to drift based of the magnetic polarity of Sun and the charge sign of the GCR. During so-called A $>$ 0 polarity cycles such as solar cycle 24, when the HMF is directed away from the Sun in the northern hemisphere, positively charged particles drift to the Earth more directly and efficiently from the polar to the equatorial regions and outwards along the HCS, while negatively charged particles drift in opposite directions, through equatorial regions, where they encounter greater scattering and gradient in the Parker-spiral solar wind~\citep{parker_passage_1965}. As a consequence, propagation time and energy loss depend on the period of the solar cycle, the polarity state, and the charge of the particle. The solar magnetic field reverses polarity at each solar maximum. In recent high-precision measurements, PAMELA and AMS-02 have reported on the charge-sign dependence of electron and positron propagation through the heliosphere down to 500 MeV~\citep{adriani_time_2016,aguilar_observation_2018}, confirming the results from the balloon-borne mission AESOP in previous decades~\citep{clem_positron_2002}.  \newline \indent
However, despite the great progress made with the advent of high precision space missions, a gap in the understanding of the cosmic ray electron and positron spectra remains below 100 MeV. In this energy range, measurements made on Earth can be compared to the unmodulated Local Interstellar Spectra (LIS), now probed by the two Voyager spacecraft, which crossed the heliopause in 2012 and 2018~\citep{stone_voyager_2013,stone_cosmic_2019}. The negative spectral index below 100 MeV has been observed since the 1960's~\citep{earl_cloud-chamber_1961,meyer_electrons_1961}, yet the mystery of its origin persists. Whereas well-established comprehensive three-dimensional numerical models have been used along with experimental data from Voyager, PAMELA, and AMS-02 to reproduce CR spectra at 1 AU~\citep{potgieter_modulation_2015,vos_new_2015,bisschoff_new_2019}, the lack of empirical knowledge in the low-rigidity regime hinders a full test of charge-sign dependent solar modulation, provided by a simultaneous measurement of particle/antiparticles species. \newline \indent

\begin{figure}[ht!]
\centering
\includegraphics[width=0.60\textwidth]{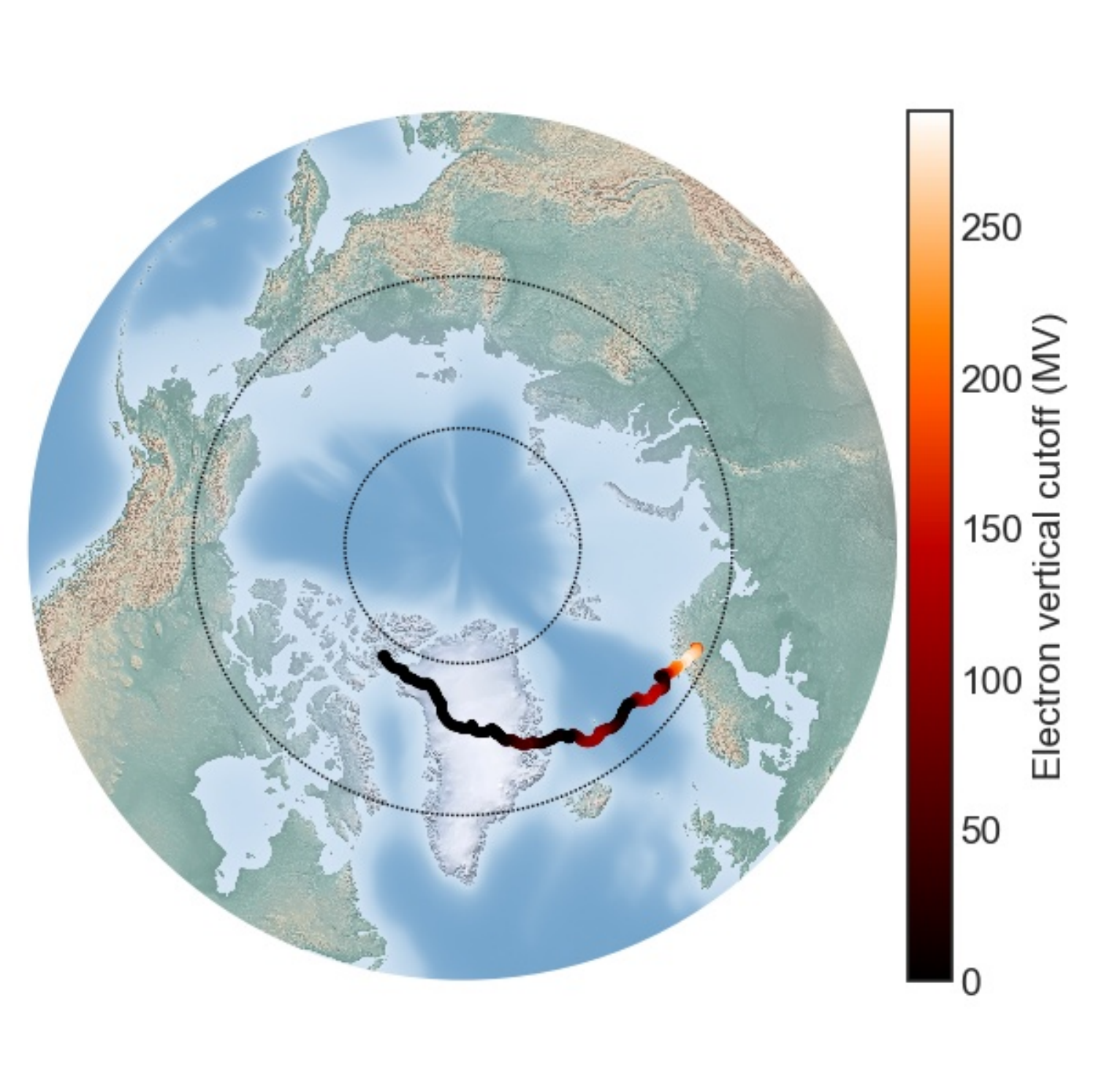}
\caption{Trajectory of the first flight. The first 90 hours of the flight surveyed latitudes where diurnal variations of the geomagnetic field are still present, as indicated by the color-coded legend. \label{fig:Flightmap}}
\end{figure}

This paper reports on observations of electrons and positrons by the AESOP-Lite balloon-borne experiment. The instrument successfully completed its first mission in May 2018, on a 5-day flight between Esrange, Sweden (67$^{\circ}$89'N) and Ellesmere Island, Canada (78$^{\circ}$40'N), on a NASA 40~million cubic feet ($\sim$1.1 million m$^{3}$), zero pressure, long duration balloon. The balloon floated at an average altitude of 135~kft ($\sim$ 41 km), which corresponds to $\sim$ 2-4~g.cm$^{-2}$ of atmospheric overburden, collecting data for roughly 133 hours.  The northerly trajectory of the payload (Fig.~\ref{fig:Flightmap}) allowed the apparatus to survey regions of low geomagnetic cutoff (below 200 MV). We present here our measurement of the primary electron and positron spectra between 20 MeV and 1 GeV, as well as the re-entrant albedo fluxes in the same energy range. We describe the detector system in \textsection\ref{sec:detector}, the data analysis in \textsection\ref{sec:selection}, and the results are presented and discussed in \textsection\ref{sec:results}.

\section{The detector system} \label{sec:detector}
\begin{figure}
\centering
\includegraphics[page=1,width=0.85\textwidth]{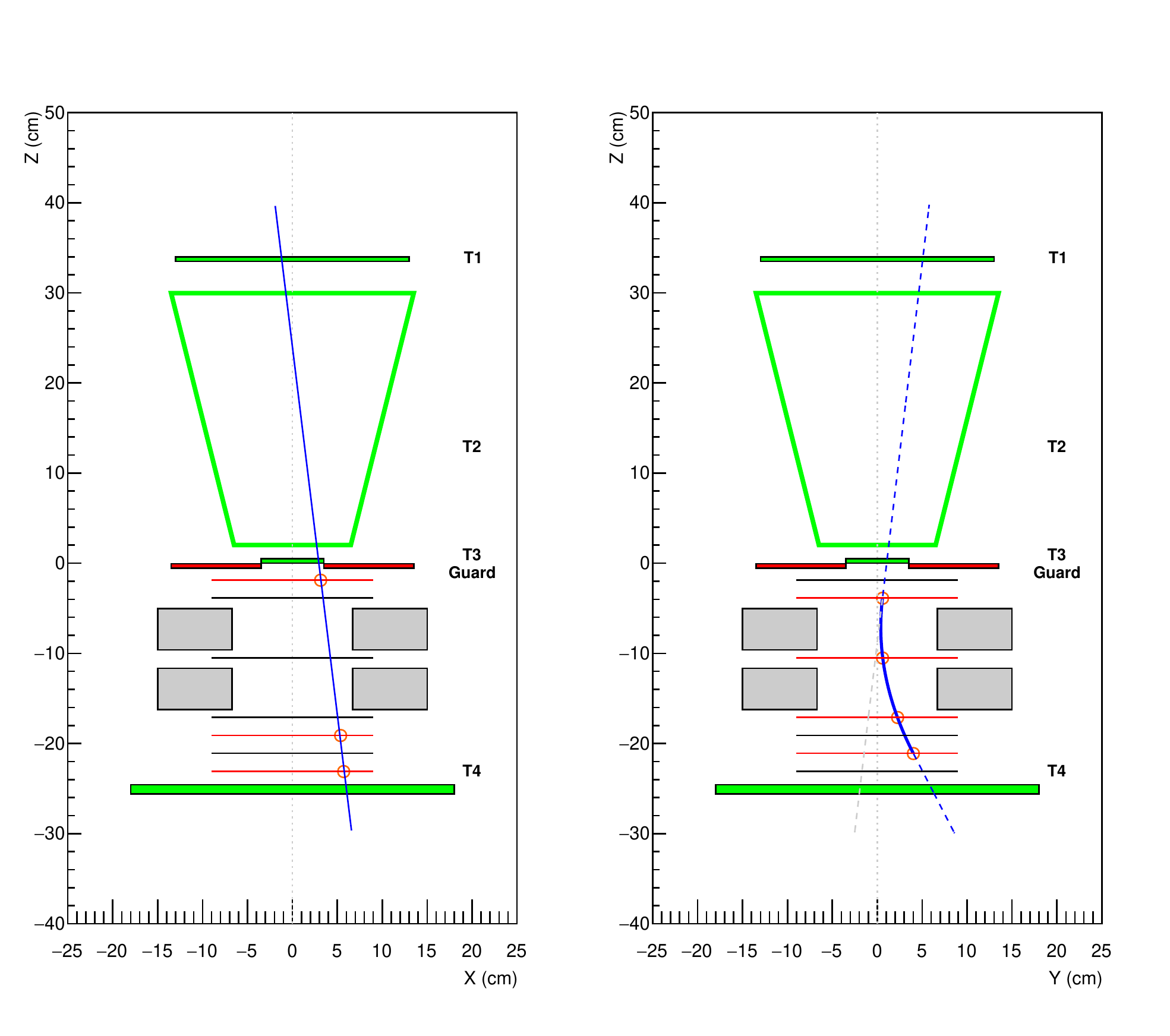} 
\caption{Cross section of the AESOP-Lite instrument as viewed from the event display software. Shown is an electron candidate recorded during the 2018 flight. The triggers T1, T2, T3 and T4 were fired (in green), whereas no signal was seen in the guard (in red). The active layers in each view (non-bending on the left, bending on the right) are shown in red.}
\label{fig:EventDisplay}
\end{figure}

The AESOP-Lite apparatus is the successor to the LEE (Low Energy Electrons) payload ~\citep{hovestadt_detector_1970}, which retired after 23 successful flights, having contributed to important measurements of low-energy electrons over many solar cycles~\citep{fulks_solar_1975,evenson_quantitative_1983,evenson_cosmic_2009}. The measurements from the final LEE flights, which occurred in 2009 and 2011 (``LEE09'' and ``LEE11''), were analyzed using the same method outlined in~\cite{fulks_solar_1975}. The results, used in this analysis, are presented for the first time in Appendix A. The LEE instrument was modified by replacing the original calorimeter with a magnetic spectrometer, making it possible to resolve charge-sign. The original entry telescope was preserved. Since the pulse height analyzers (PHA) and front-end electronics of each counter-photomultiplier tube (PMT) system were used in past LEE flights, we are provided with a mean of cross-calibrating the absolute electron fluxes with previous measurements. Fig.~\ref{fig:EventDisplay} presents a diagram of the instrument. 

\subsection{The entry telescope} \label{subsec:telescope}
Detectors T1, T2, T3, T4, and the guard are all read out by PMTs. The entry telescope consists of three NE 102 A plastic scintillators (T1, T3 and Guard) and a gas Cherenkov detector (T2). T3 identifies singly charged relativistic particles, while T2 serves as a hadron discriminator. The Cherenkov counter is filled with C$_{3}$F$_{8}$ gas to an absolute pressure of 1.8 atm, rejecting all particles with $\gamma = $E/mc$^{2}\leq~15.7$ (corresponding to an electron energy of approximately 8 MeV). Furthermore, T2 is only sensitive to downward-moving particles, thus removing upward-going splash albedo present in the atmosphere. The  T1--T2--T3 coincident signal  is used in flight as the main trigger for the tracker system. It also defines the geometric acceptance of the instrument, $\sim$ 18 cm$^{2}$sr, without taking into account the important loss in fiducial volume incurred by the presence of the magnetic field.
The guard counter (G), strictly used offline in anti-coincidence, serves to flag particles produced by showers inside the apparatus, whereas the plastic scintillator T4, placed at the very bottom, selects particles that have exited the bottom of the instrument.
\subsection{Tracking spectrometer} \label{subsec:tracker}
The tracking system consists of seven planes of silicon strip detectors (SSD) and a Halbach ring dipole magnet~\citep{halbach_design_1980}. The average field is 0.3~T, though its known non-uniformity must be accounted for. The SSDs are arranged in an xy-configuration, with 4 layers in the bending plane to measure the particle deflection, and 3 layers to view their trajectory in the non-bending plane.  The magnet design allows the placement of a tracker in the bending-view at the center of the field. The silicon wafers were custom designed and manufactured for the Large Area  Telescope (LAT) of the NASA Fermi mission~\citep{atwood_design_2007}. Each SSD is a 8.95$\times$8.95~cm$^2$, 400~$\mathrm{\mu}$m thick single-sided detector, with strip pitch 228~$\mathrm{\mu}$m and spatial resolution 66~$\mathrm{\mu}$m ($228/\sqrt{12}$). \newline \indent

In flight, coincidences T1--T2--T3 and T1--T2--T4 were alternately used as an online trigger in flight (the ``GO'' signal). The  tracker system self-triggers with a logical OR of two triggers:  one from the bending plane, the other from the non-bending, requiring in each view a coincidence of at least one strip hit in each of the top 3 layers. The data are stored in each board until a ``GO'' signal is received. If a ``GO'' signal fails to arrive within 5~$\mathrm{\mu}$s the data are discarded. 

\section{Analysis} \label{sec:selection}
To identify electrons and positrons from the total data sample collected in flight, a set of selection criteria is established. The level of background atmospheric particles is then estimated in order to derive the primary spectra. This analysis work relies on two Monte Carlo (MC) simulations performed with the FLUKA 2011 software~\citep{ferrari_fluka:_2005,bohlen_fluka_2014}.

The first simulation studies the instrument's response to fluxes of electrons, positrons, and protons. Due to the acceptance of the entry telescope, particles are generated with an incident angle $\theta<40^{\circ}$. The simulation includes the magnetic field map provided by the magnet manufacturer.
The efficiency of each selection is calculated using the MC results and/or the flight data when possible. 

In addition, we have completed MC simulations of the atmospheric air shower induced by incident fluxes of GCR protons and alpha particles. The FLUKA code interfaces with the DPMJET-3 hadronic model~\cite{roesler_monte_2001} for GCR nuclei above 5 GeV/nucleon.

\subsection{Particle Identification}

\subsubsection{Tracking} \label{subsec:Tracking}
The spectrometer measures the rigidity of a particle that has successfully passed the online trigger requirement.
The event is first processed with a pattern recognition routine, which selects hits belonging to the same track.
We impose a set of conditions on the fitted track to obtain a reliable reconstruction:
\begin{itemize}
\item We restrict the allowed range of hit positions in the first three tracking layers to eliminate events that have scattered near or in the magnet walls, or produced electromagnetic showers in the upper-half of the spectrometer. All dimensions are given as measured from the center of the layer:
\begin{itemize}
\item In L$_{0}$: $\mid x \mid$ $<$ 4.0 cm 
\item In L$_{1}$: $\mid y \mid$ $<$ 6.0 cm 
\item In L$_{2}$: $\mid y \mid$ $<$ 6.25 cm 
\end{itemize}
\item At least 6 (of 7) tracking layers must have a hit, with a maximum of 9 hits. We demand that all 4 layers in the bending view record a hit, and that at least 2 (of 3) layers in the non-bending view do so. This condition eliminates multi-track and $\delta$ -rays events. 
\end{itemize}

This selection helps to exclude events that have interacted in the tracking volume, or crossed a region of weak magnetic field.
In the non-bending view, the algorithm fits all possible lines between hits in the top-most and bottom-most layers, and chooses the track that minimizes the $ \chi^{2} $ to derive the dip angle $\theta_{NB}$ . In the bending view, a second order polynomial function is fit to all possible configurations of hits in the four layers, and the best fit is chosen. The radius of curvature is calculated for all 4 bending planes, and the mean is taken to infer the transverse momentum  $p_{T}$.

Fig.~\ref{fig:EventDisplay} illustrates the parabola and straight line fits as seen in solid blue lines in the event display. The dashed lines in the bending plane indicate the incoming and outgoing directions of the particle, assuming no scattering or interaction has occurred in the detector.  

Results from the pattern recognition are then used to initialize a fourth-order Runge-Kutta numerical integration: this method can follow a charge-particle’s changing momentum through an arbitrarily changing magnetic field, as long as multiple scattering in the silicon can be neglected; it serves as the final track fit method. A track is characterized by five parameters: $x$, $y$, the position of the particle at the first spectrometer layer, $cx$, $cy$, its directional cosines at at that point, and transverse momentum $p_{T}$. The track parameters  are iteratively adjusted by a Nelder-Mead "simplex" algorithm~\citep{nelder_simplex_1965} until a minimum is found with respect to the numerically integrated trajectory.

The resulting maximum detectable rigidity at 3$\sigma$ is $\sim$ 900~MV, with a reconstruction resolution of $ \sim $ 9\% at 20~MeV/c and $ \sim $ 18\% at 300~MeV/c, based on MC results obtained from fitting the 1/p$_{reco}$ distribution to a Gaussian function. The mean and resolution of the reconstruction at each momentum are derived from the mean and the width of the fitted Gaussian. The validity of the charge-sign capabilities of the spectrometer and the momentum reconstruction was verified using ground-level muons.

In addition to the tracking requirements listed above, a good quality of fit is demanded, as given by the value of the mean square deviation of the fit.

\subsubsection{Scintillator and Cherenkov selection} \label{subsec:particleID}
The coincidence signal T1-T2-T3 is used as the main trigger in flight. Offline, we add the guard in anti-coincidence, as a way to remove events that have produced an electromagnetic shower above the spectrometer. A sequence of cuts is applied to identify electrons and positrons from the raw data sample. The scintillators T1 and T3 efficiently detect particles with $\mid Z \mid$ = 1, whereas the main purpose of the Cherenkov detector T2 is to discriminate against the high background of protons and muons present in the atmosphere. Finally, a particle must also cross the bottom-most scintillator T4 to be considered an electron or positron candidate.

\begin{figure}[ht!]
\centering
\includegraphics[width=0.49\textwidth]{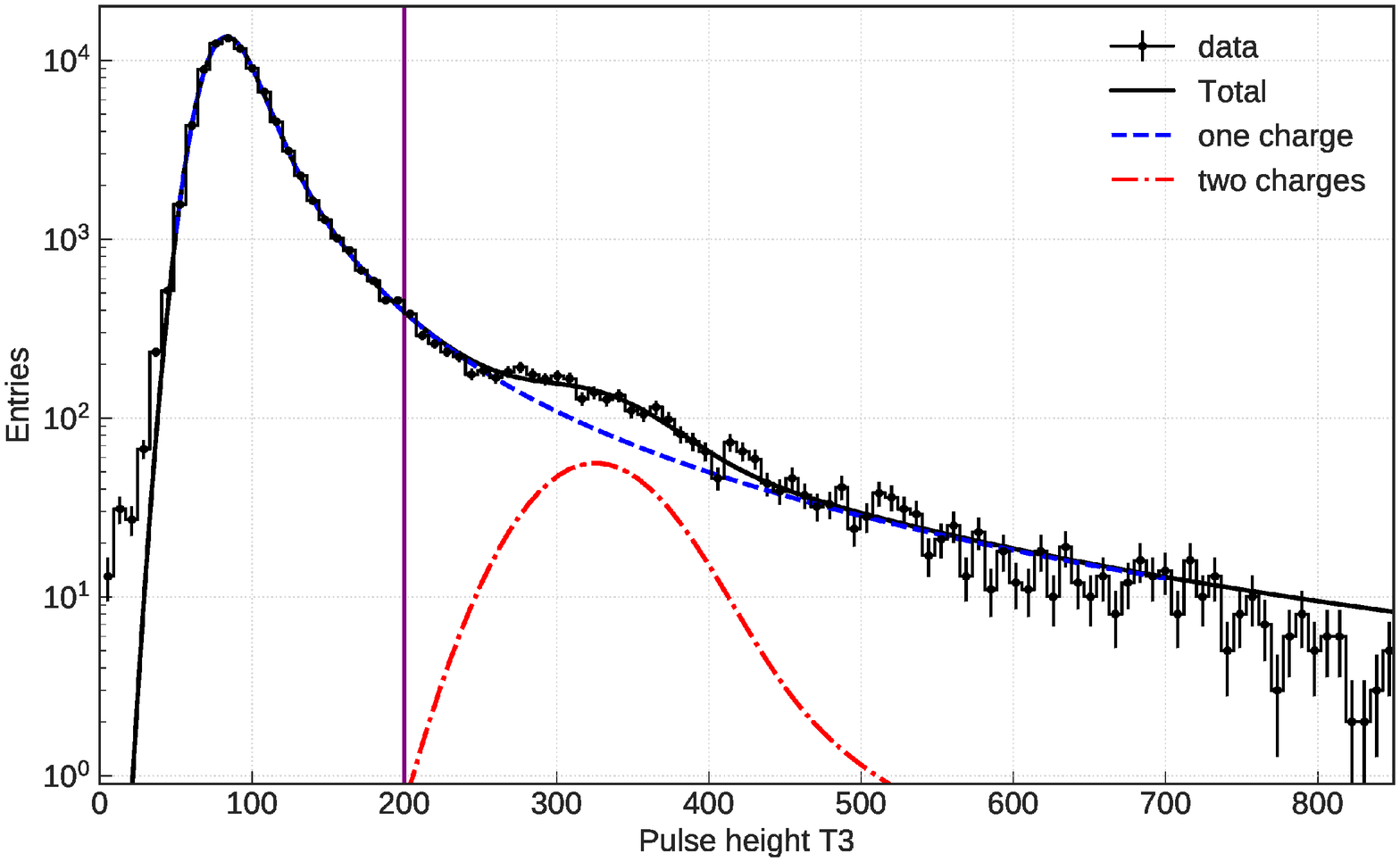} 
\includegraphics[width=0.49\textwidth]{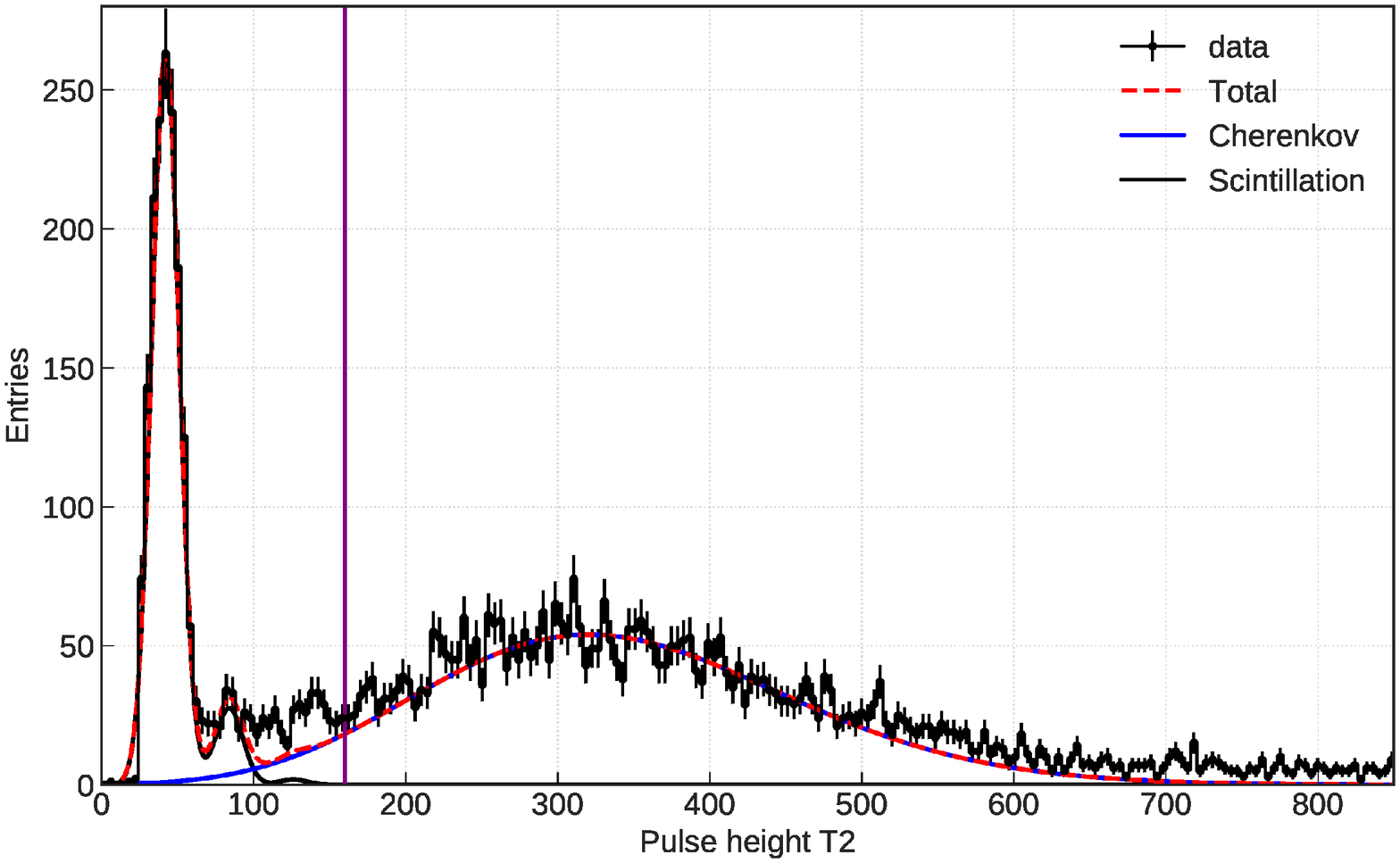} 
\caption{Left: PHA distribution of the scintillator T3 during the 2018 flight. The vertical line shows the upper-limit cut. The global fit (sum of two Laundau-Gaussian) is shown in black, the contribution from $Z$=1 particles is shown with the dashed blue line, and that from $Z$=2 particle with the dotted dashed red line. Right: PHA distribution of T2 during the 2018 flight for tracks with a reconstructed momentum between 271 and 419 MeV/c. The purple vertical line is the offline lower threshold applied to the T2 signal (See the text for more details).
\label{fig:T2T3signal}}
\end{figure}

A cut on the signal in T3 is applied to discriminate against alpha particles, as the energy deposited grows proportionally to $Z^{2}$, as shown in Fig.~\ref{fig:T2T3signal} (left). Alpha particles with a kinetic energy below $\sim$ 54 GeV will not produce a Cherenkov signal in T2, but they can produce a signal through scintillation. We fit the total distribution of the signal in T3 for the reconstructed tracks passing all the other selection criteria. The fit function is the sum of two Landau-Gaussian functions, one for the signal from singly charge particles and one for that from double charge particles. The fit values of the Landau MPV (Most Probable Value) are $\sim$ 83~ADC and $\sim$ 327~ADC for the alphas, about 4 times the value of a $Z$=1 particle, as expected. The upper-limit cut of 200~ADC counts was chosen to remove the contamination from alphas. The residual alpha contamination is negligible.

In T2, the Lorentz threshold $\gamma$ is set to a value $\gamma$ = 15.7, such that protons with kinetic energy below 13.8~GeV and muons below 1.5~GeV do not trigger the detector. The online threshold applied on the PMT signal of T2 is set below the 1 photo-electron (PE) signal. Fig.~\ref{fig:T2T3signal} (right) shows the T2 signal for all tracks with reconstructed momentum between 271 and 419~MeV/c after the full sets of selection was applied. Two contributions can be observed. The large peak at 1 PE ($\sim42$~ADC) and the smaller peaks at 2 and 3 PE are attributed to scintillation produced by low energy protons in the C$_{3}$F$_{8}$ gas, fit with a function that is the sum of 3 Gaussians (black line). The signal above the peaks is associated with the Cherenkov light produced by the electrons and positrons, fit to a Poisson distribution (blue line). For this range of reconstructed momentum, an average signal is estimated to be 8 PE. The total fit, which is the contribution of the scintillation and the Cherenkov signal is shown in red.

We apply a cut on the PHA value of T2 at an ADC count value of 160 to remove the contamination (which is negligible after this cut). With this selection, the AESOP-Lite and LEE count rates agree within 5\% during the ascent phase of the flight, when the temporal variation of cosmic rays flux is less relevant than at float altitude. For each bin in reconstructed momentum, the efficiency losses due to this selection are calculated assuming a Poisson distribution for the number of Cherenkov PE. The signal losses are $\sim$11\% at 25~MeV/c, $\sim$7\% at 45~MeV/c, and $\sim$5-6\% above.  The total numbers of reconstructed events per bin in momentum are corrected accordingly.

\subsubsection{Detection efficiency} \label{subsec:efficiency}
\begin{figure}[ht!]
\centering
\includegraphics[width=0.5\textwidth]{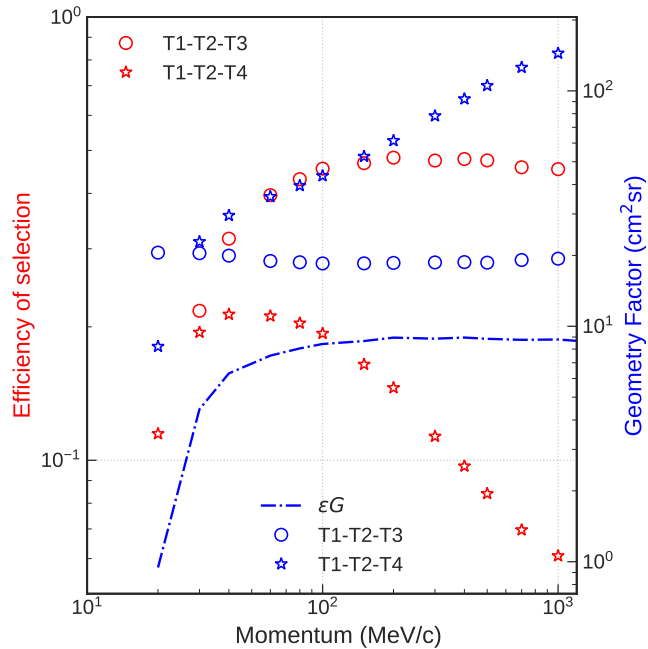}
\caption{The selection efficiencies $\epsilon_{sel}$ and geometry factor for trigger modes T1-T2-T3 and T1-T2-T4 for electrons.}
\label{fig:eff}
\end{figure}

The total detection efficiency of the final event sample $\epsilon$ can be divided in two components: the trigger efficiency $\epsilon_{trigger}$ and the particle selection efficiency $\epsilon_{sel}$. The former is defined as $\epsilon_{trigger} = \epsilon_{GO} \times \epsilon_{tkrtrig}$, with $\epsilon_{GO}$ the efficiency of the main trigger (T1-T2-T3 or T1-T2-T4), estimated to be $\sim$99\%, and $\epsilon_{tkrtrig}$ the tracking system trigger efficiency. The latter is the efficiency of the tracker trigger, which is an ``OR'' of the trigger signal in the bending and non-bending planes. To evaluate it, we have used a combined analysis of MC and flight data to estimate the inefficiencies each view's trigger, which were above 50\% in the bending plane in flight, due to faulty connections between trackers, and $\sim$ 80\% in the non-bending view. In total, we estimate $\epsilon_{tkrtrig}$ to be 94\%. \\
The efficiency $\epsilon_{sel}$ of the particle's selections of Sec. \textsection\ref{subsec:particleID} is calculated, normalized to the number of entries for the two online triggers in flight, $\epsilon_{T_{1}T_{2}T_{3}}$ and $\epsilon_{T_{1}T_{2}T_{4}}$, using MC simulations only. They are shown in red in Fig.~\ref{fig:eff}. The efficiency of the selection on the PHA value of T2, $\epsilon_{T2}$, is excluded from what is presented in Fig.~\ref{fig:eff}, as  $\epsilon_{T2}$ is derived from flight data. 

The two geometry factors are shown in blue in Fig.~\ref{fig:eff}. The total efficiency times geometry factor $\epsilon G$, expressed in cm$^{2}$sr, is represented by the dashed line. It is identical for both trigger configurations, as the increased geometry factor of T1-T2-T4 compensates for the smaller selection efficiency.  The small differences in the geometry factor and selection efficiency that exist between electrons and positrons (especially at low energy) have been studied and taken into account in the analysis.

\subsection{Interplanetary fluxes} \label{subsec:FluxesTOA}
Once these selections have been completed, several steps are needed to extract the interplanetary, or primary, electron and positron differential flux spectra from a sample of event candidates. We describe our method in this section. 

\subsubsection{Time selection}
Although the northerly trajectory of the payload allowed us to survey latitudes of low rigidity cutoff $E_{c}$ (below 200MV), the diurnal variations between geomagnetic day and night were still present~\citep{jokipii_diurnal_1967} until the balloon reached latitude $\sim$ 70$^{\circ}$N after 90 hours of flight. The particle rate rises during geomagnetic day when upward-going secondary ``splash'' albedo particles produced in the interaction of primary cosmic rays with atmospheric nuclei, lacking the energy to escape the geomagnetic field lines, spiral along them to reach their conjugate point, at the opposite latitude. These downward-going electrons are then called ``re-entrant'' albedo particles, overwhelming the primary electron signal by this trapped secondary component~\citep{israel_cosmic-ray_1969,verma_measurement_1967}. During geomagnetic night, however, as the field lines extend, the geomagnetic cutoff becomes essentially null: ``splash'' albedo particles can safely escape, and primary cosmic-ray particles of all energies are able to enter the atmosphere. \\

\begin{figure}[ht!]
\centering
\includegraphics[width=0.75\textwidth]{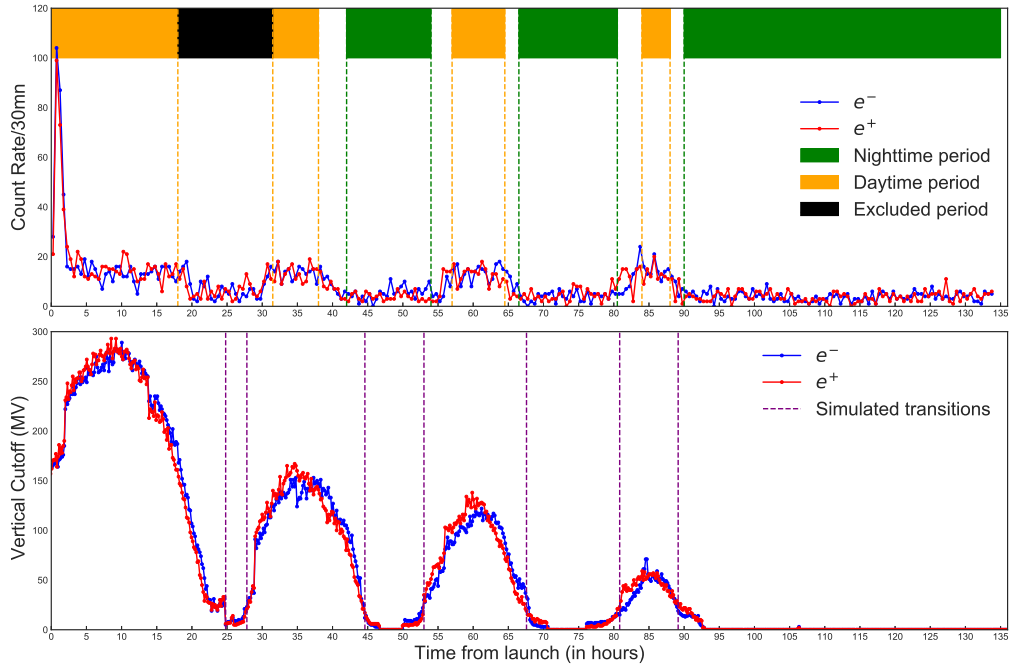}
\caption{Top: Time series of reconstructed electrons and positrons at the lowest momentum bin (20-30~MeV/c). The diurnal variations between geomagnetic day and night are clearly visible. Bottom: Simulated variation of the vertical geomagnetic cutoff.}
\label{fig:timeserie}
\end{figure}

The instrument recorded $\sim$ 70 hours of data during geomagnetic night. To separate our events in periods of ``daytime'' (DT) and ``nighttime'' (NT), we compare the time series of reconstructed electrons and positrons at the lowest energy bin with calculated variation of the vertical geomagnetic cutoff (Fig.~\ref{fig:timeserie}), based on measurements of the Kp index~\citep{bartels1949standardized}, indicating the level of geomagnetic disturbances at the time of flight. To compute the geomagnetic cutoff, we use a code ~\citep{lin_electron_1995} that calculates the trajectory of particles based on the IGRF (International Geomagnetic Reference Field) for the internal geomagnetic field and the Tsyganenko model of the magnetosphere ~\citep{tsyganenko_global_1987}. Apparent jumps in the cutoff calculation occur because the Kp index is only defined in three hours interval. Daytime and nighttime time zones are selected when the transitions in the flight data and the simulation agree with one another. When they do not, the region is excluded from the analysis (the black hatched section in Fig.~\ref{fig:timeserie}). \\

Within daytime and nighttime sets, we further section the event sample into 23 time bins $\Delta T$, ranging from 15 minutes intervals, to capture the ascent, to $\sim$ 9 hours at float altitude. For each average atmospheric depth $d$ in the time bin, the spectra of electrons and positrons as a function of reconstructed momentum are calculated. These spectra are then ready to be unfolded.

\subsubsection{Electron and Positron fluxes at the Top of Payload} \label{subsec:FluxesToP}

Before reaching the spectrometer, a minimum ionizing particle will lose about 4 MeV in the shell and scintillators of the entry telescope. To contend with the biases, inefficiencies, and finite resolution of the energy reconstruction, we simulate a response matrix that encodes the smearing of the desired true quantity into the measured observable. A deconvolution, called unfolding, is performed to estimate the true variable. An iterative statistical procedure, based on Bayes' theorem, was developed by \cite{dagostini_multidimensional_1995}. For this work, we have used the Python package \textit{PyUnfold}, which was developed for the HAWC cosmic--ray experiment and implements the aforementioned unfolding algorithm~\citep{bourbeau_pyunfold_2018}.

\begin{figure}[ht!]
\centering
\includegraphics[width=0.505\textwidth]{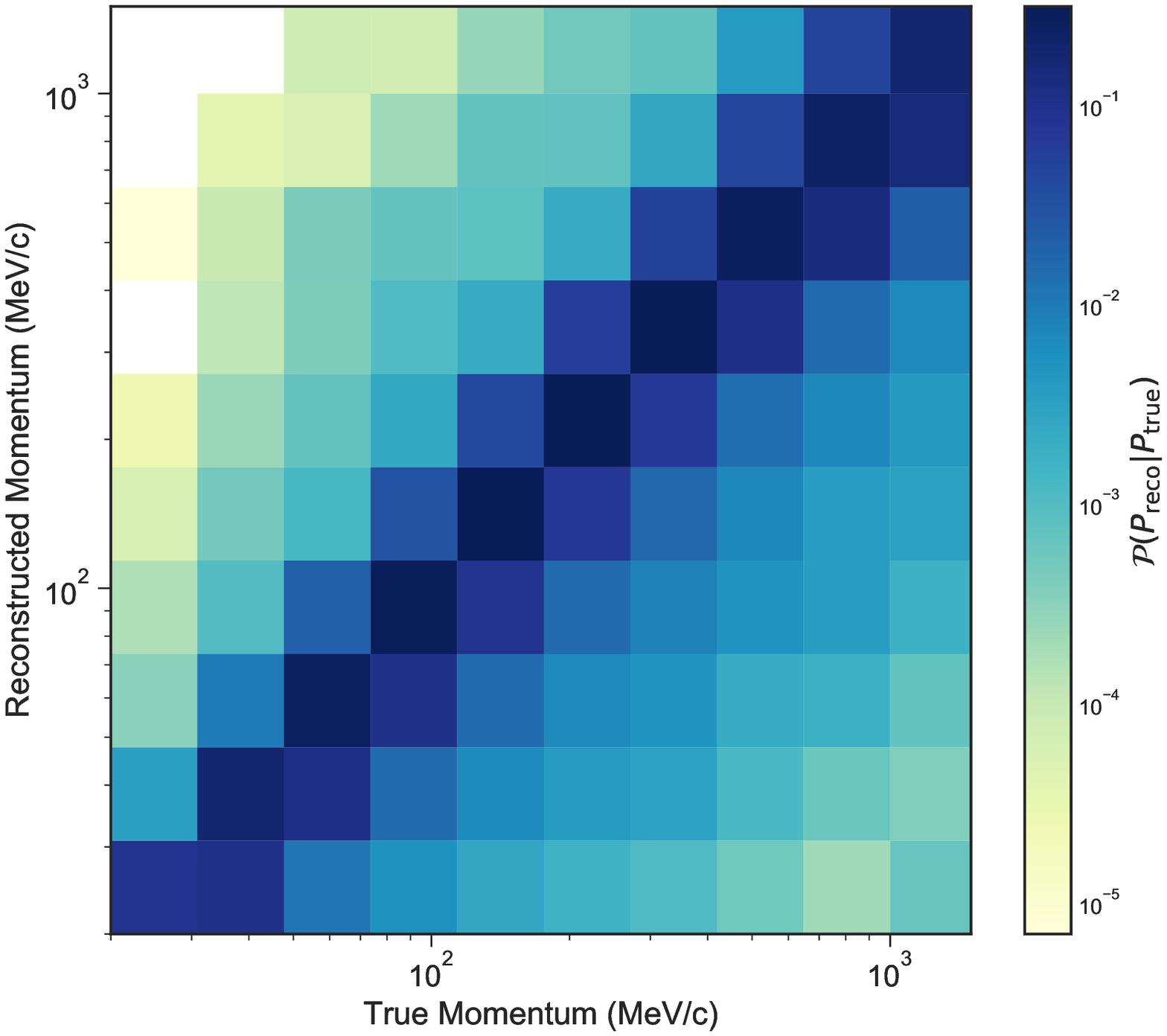}
\includegraphics[width=0.45\textwidth]{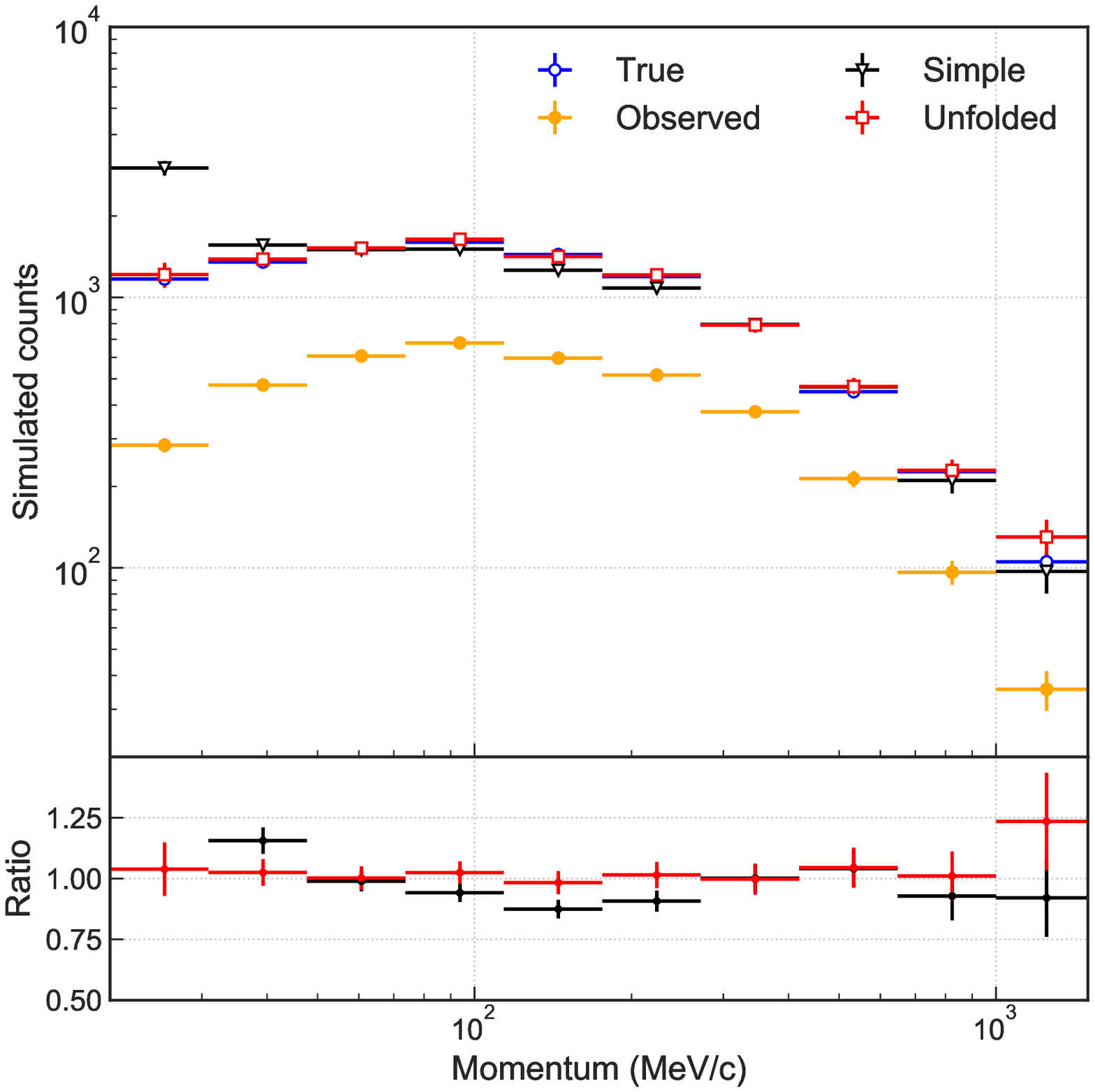}
\caption{Left: Example of the normalized response matrix weighted for one of the time bin at float altitude. The true and reconstructed momenta are plot on the x and y axis, respectively. Right: Simulated true, observed and unfolded counts as a function of momentum expected during the same time bin. The bottom shows the ratios between the reconstructed counts and the expected true counts with (red) and without (black) unfolding. The presented results are for electrons and are based on Monte-Carlo only.}
\label{fig:Response}
\end{figure}

We generate a set of MC electrons in the energy range 10-1500 MeV, following a $P^{-2}$ power-law  distribution in momentum $P$, and select particles that have passed the full flight criteria. 

The simulated response matrix, whose elements represent the probability for an electron of momentum $P_{\mathrm{true}}$ to be reconstructed with a momentum $P_{\mathrm{reco}}$, is shown in Fig.~\ref{fig:Response} (left). We observe that its diagonal elements dominate, while deviations from the diagonal represent the bias and resolution of the reconstruction. Prior to being used on the flight data, the method was tested with an independent distribution of simulated electrons, to be unfolded using the response matrix. For each time bin, the response matrix is weighted with the expected background spectrum at that given altitude.
For instance, Fig.~\ref{fig:Response} (right) shows the comparison of the simulated true counts (blue), the observed counts (orange), the unfolded counts (red) and the counts determined without unfolding (black) for one time bin.
The unfolding procedure was found to improve the accuracy of the reconstructed distribution by as much as 20 \% for lower energy bins, compared to a reconstruction \textit{sans} unfolding, as illustrated by the bottom panel of Fig.~\ref{fig:Response} (right).

\begin{figure}[ht!]
\centering
\includegraphics[width=0.45\textwidth]{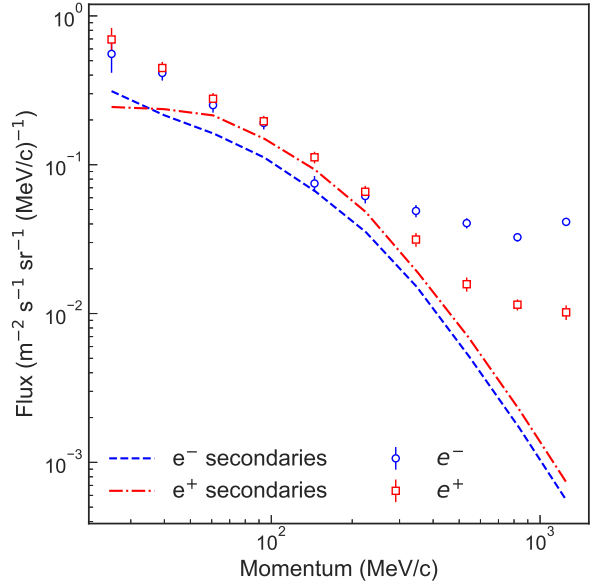}
\caption{Unfolded electron (blue) and positron (red) spectra at the top of the payload and the simulated atmospheric secondary electrons and positrons (dashed lines) at 2.5 g cm$^{-2}$ of residual atmosphere.}
\label{fig:UnfoldedSpectraTOP}
\end{figure}

For each time bin the data are unfolded by normalizing the response matrix to the calculated efficiency $\epsilon_{sel}$ of the final selection. The unfolding procedure is carried through to yield the corrected count $N_{e^{-},e^{+}}$. The differential flux $\Phi_{e^{-},e^{+}}(P)$ can then be derived:

\begin{equation}
\Phi_{e^{-},e^{+}}(P) = \frac{N_{e^{-},e^{+}}}{\Delta T \times \epsilon_{trigger} \times G(P) \times \Delta P},
\label{flux}
\end{equation}

with $\Phi_{e^{-},e^{+}}(P)$ in m$^{-2}$sr$^{-1}$s$^{-1}$(MeV/c)$^{-1}$, $\Delta T$ the time interval in s, and $\epsilon_{trigger}$ the trigger efficiency described in Sec. \ref{subsec:efficiency}. $G(P)$ is the geometry factor in m$^{2}$sr, and $\Delta P$ is the width of the momentum bin in MeV/c. An example of these fluxes is shown in Fig.~\ref{fig:UnfoldedSpectraTOP}, in which the edge bin 1--1.5 GeV/c is unfolded but not used.

\begin{figure}[ht!]
\centering
\includegraphics[width=0.32\textwidth]{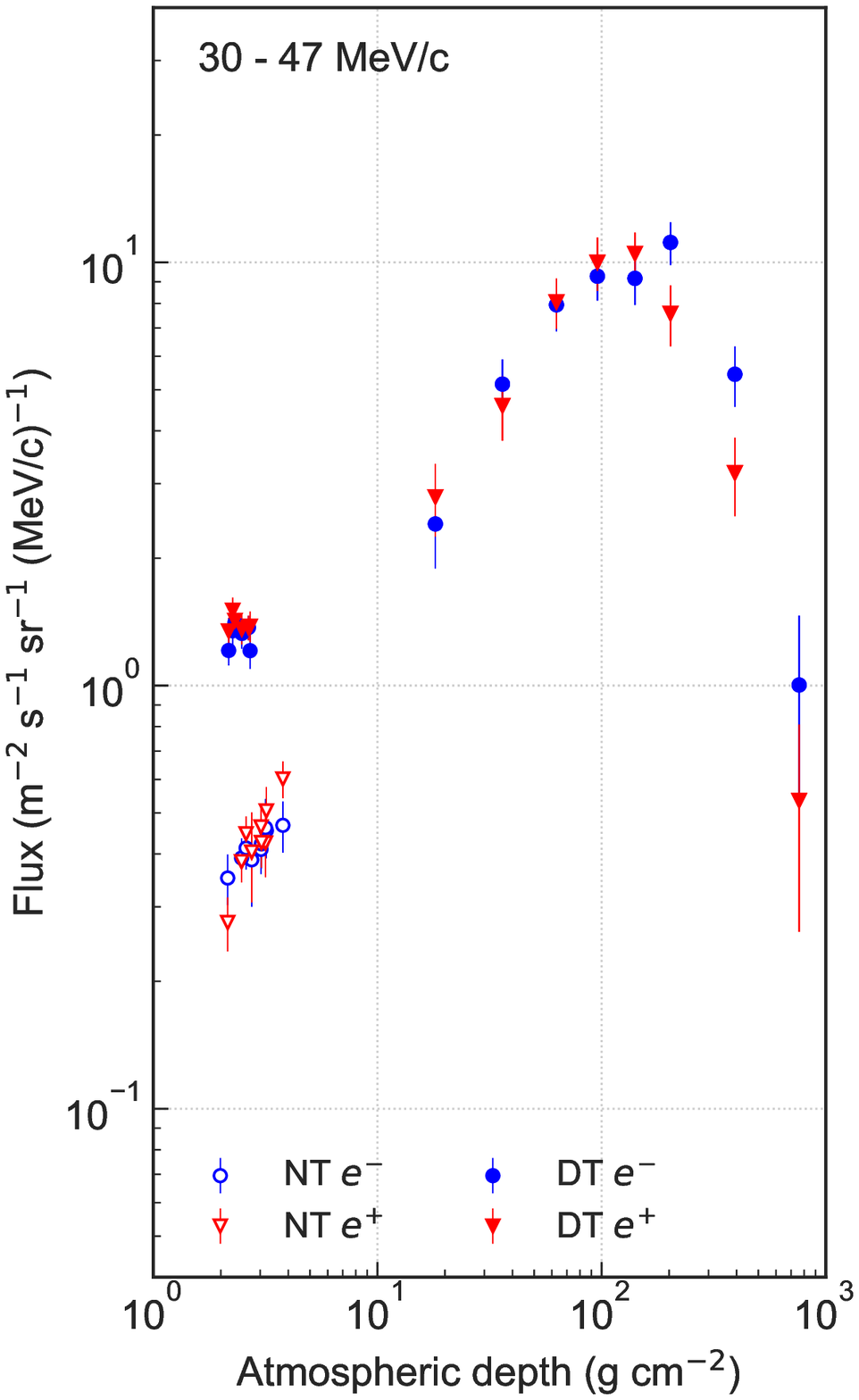}
\includegraphics[width=0.32\textwidth]{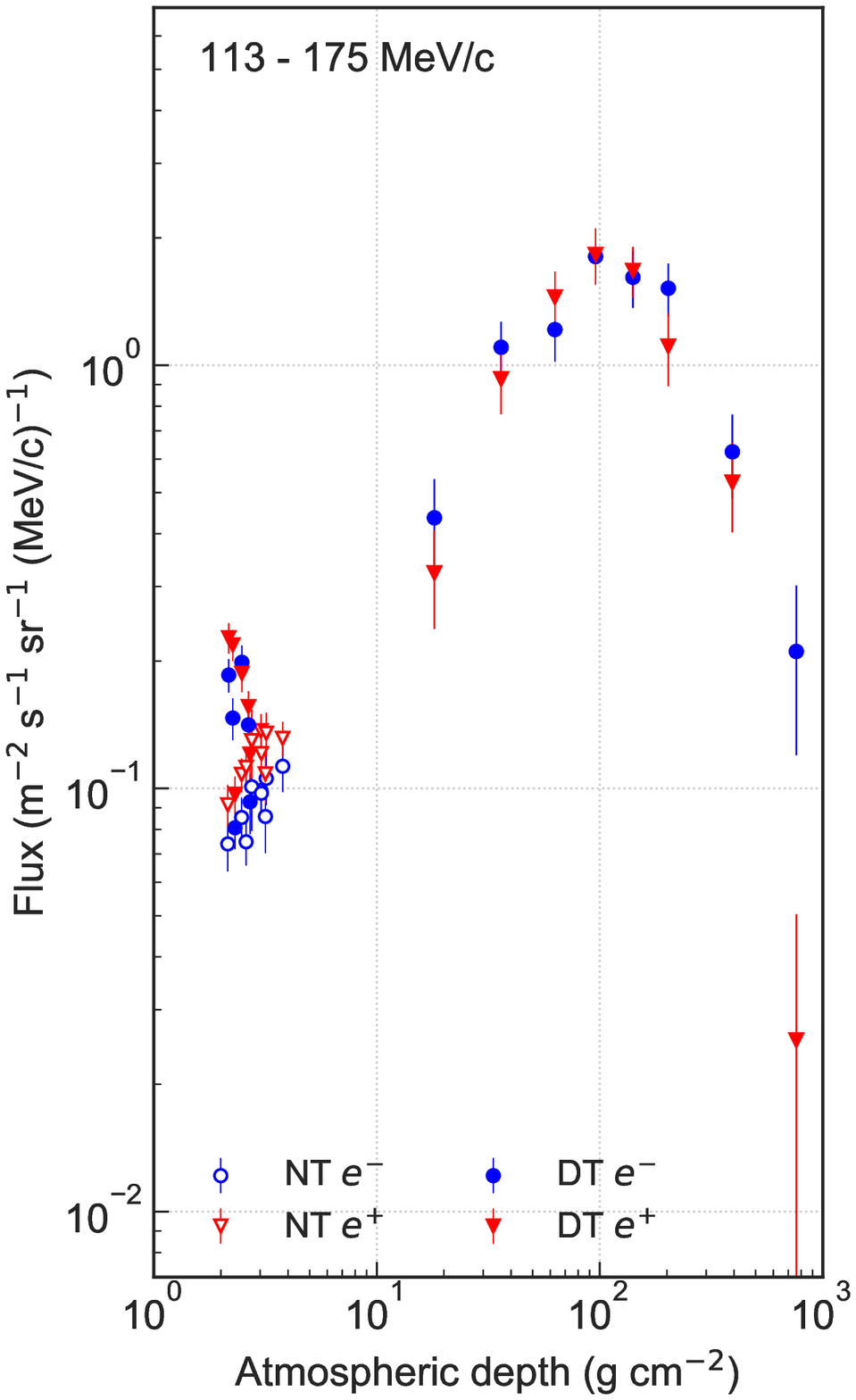}
\includegraphics[width=0.32\textwidth]{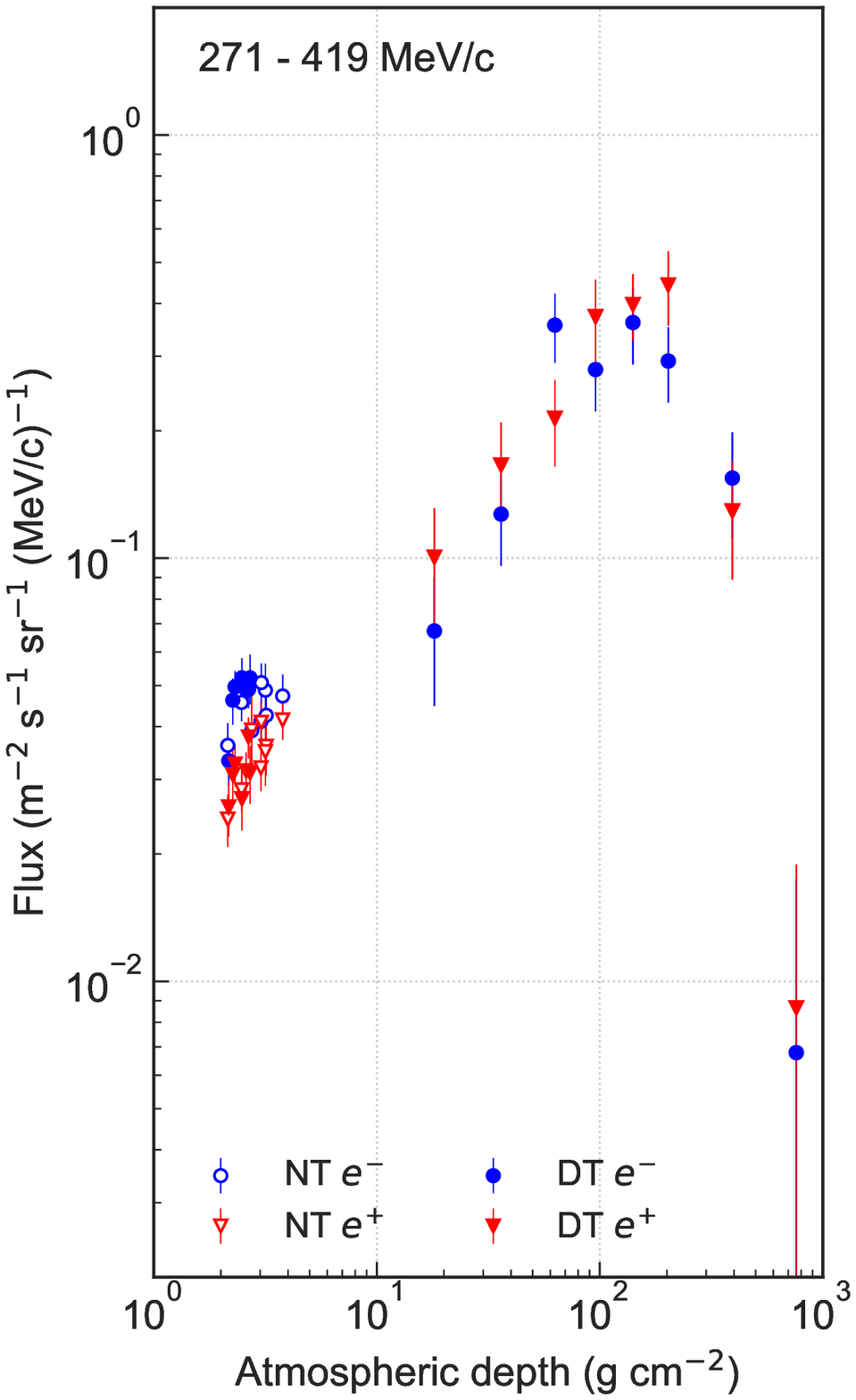}
\caption{Daytime (DT) and nighttime (NT) growth curves for 3 ranges of momentum.  In the range 30-47 MeV/c, the flux of re-entrant albedo particles at float altitudes is clearly visible during daytime (left), whereas the difference between nighttime and daytime is hardly visible for the range 271-419 MeV/c (right). This is in agreement with our estimation that the vertical cutoff during the flight did not exceed 300 MV. The middle panel shows the daytime and nighttime points in the geomagnetic transition zone.}
\label{fig:GCdata}
\end{figure}

Once all time-separated spectra are unfolded, the data set is organized by momentum bins chosen in uniform logarithmic space between 20~MeV/c to 1~GeV/c. We then produce growth curves for each bin, that is, a profile of the flux of particles as a function of the atmospheric depth. The first 2.5 hours are used to obtain the points during the ascent, where the low energy cosmic ray electron and positron signal is assumed to be atmospheric secondaries. 
Fig.~\ref{fig:GCdata} presents the growth curves for 3 ranges of energy: 30--47 MeV/c (left), 113--175 MeV/c (center), and 271--419 MeV/c (right). 
The ascent and the first 17 hours of the flight occurred during geomagnetic daytime. The corresponding data are presented with filled markers in the figure. The nighttime data set is shown with open markers. 
As seen in the bottom panel of Fig.~\ref{fig:timeserie}, the vertical geomagnetic cutoff varies within the range 0--300~MV during the first daytime period such that we expect to observe a much larger contribution of re-entrant albedo secondary particles at low energy (below the cutoff) than at higher energies closer to the cutoff ($\sim$ 300~MV). This is indeed the case in the data presented in the left and right panels Fig.~\ref{fig:GCdata}. The bin 113--175 MeV/c represents the transition region of the geomagnetic variation: since the ascent occurred at $\sim$ 160~MV during a transition from nighttime to daytime, the daytime points represent a mixture of trapped albedo and primary cosmic rays, hence the vertical spread of the filled markers.

From this point onward in the analysis, we express all spectra in terms of kinetic energy (in MeV) instead of momentum (in MeV/c), to follow the convention used in other cosmic ray experiments.

\subsection{Flux at the top of the atmosphere} \label{subsec:BackPropagation}

For each energy bins, we distinguish three separate contributions to the flux measured at depth $d$, all derived from MC atmospheric simulations:
\begin{itemize}
\item The ``\textit{primaries}'': primary electrons and positrons that remained in the same energy bins at the top of payload (ToP) as they belonged to at the top of the atmosphere (ToA). This contribution is normalized to  a flux of 1 particle m$^{-2}$sr$^{-1}$s$^{-1}$MeV$^{-1}$.
\item The ``\textit{secondaries}'': secondary background contribution from the interactions of GCR nuclei, mostly of H and He, with the atmosphere. .
\item The ``\textit{spillover}'': the contribution of primary electrons and positrons that belonged to a higher energy bin at ToA than that they populate at ToP.
\end{itemize}

\subsubsection{Atmospheric simulations} \label{subsec:secondaries}

During the 5 days of data taking, the balloon floated at an atmospheric depth between 2 and 4~g.cm$^{-2}$. The dashed curves shown in Fig.~\ref{fig:UnfoldedSpectraTOP} are the predicted electron and positron spectra, at float altitude, of the background electrons and positrons produced in the spallation of GCR with the nuclei of the residual atmosphere. These interactions produce short-lived mesons, such as pions and kaons, which in turn decay into electrons and positrons.\\ 

To estimate this background, we implement an atmospheric simulation using a 3D profile of the atmosphere at Esrange, Sweden taken from the day of launch, following the method of~\cite{mangeard_monte_2016}.
The air shower development induced by H and He, as well as primary electrons and positrons, is simulated. Secondary particles fluxes are extracted at 19 atmospheric depths from 998 to 0.87~g.cm$^{-2}$. 

The ``\textit{secondaries}'' contribution is estimated by normalizing the simulated spectra to H and He local interstellar fluxes derived by~\cite{ghelfi_non-parametric_2016,ghelfi_non-parametric_2017}. We apply a force-field approximation of the solar modulation~\citep{gleeson_solar_1968}, using a parameter $\phi$ = 438 $\pm$ 50~MV as calculated from neutron monitor data taken at the time of the flight, using the method of~\cite{ghelfi_neutron_2017}. The heavy nuclei are assumed to produce showers similar to those from He, and are taken into account by applying a scale factor $F_{hn}=1.445$ to the He spectrum, as done in~\cite{ghelfi_non-parametric_2016}. 

We fit a 7$^{\mathrm{th}}$ degree log polynomial function to the MC results of electrons and positrons produced by protons and alpha particles. At the Regener-Pfotzer maximum ($\sim$ 100~g.cm$^{-2}$)~\citep{regener_intensity_1934}, where the flux of electrons is highly dominated by atmospheric secondaries, the MC agrees with the data by within less than $\sim$ 10\% at lower energies, while the MC systematically overestimates the signal above 200 MeV by as much as $\sim$ 40\% (see Table~\ref{tab:bpar}). This general trend of $b$ with energy is observed for all the hypothesis used in the systematic analysis, although the value of $b$ increases depending on the chosen LIS at the top of the atmosphere. 

In their propagation from the top of the atmosphere to the top of the payload, electrons and positrons experience ionization and bremsstrahlung losses, which give rise to bin migration. The simulation of the air shower development induced by primary electrons and positrons provides the ``\textit{primaries}'' and  ``\textit{spillover}'' contributions. Past analyses of balloon-borne cosmic ray data have used empirical tables of energy losses of electrons and positrons from~\cite{berger_tables_1964}, or solved the theoretical coupled cascade equations describing the propagation of electrons, positrons, and secondary gamma rays~\citep{boezio_cosmic-ray_2000}. \newline

\subsubsection{Fit method}
\begin{figure}
\centering
\includegraphics[page=14,width=0.45\textwidth]{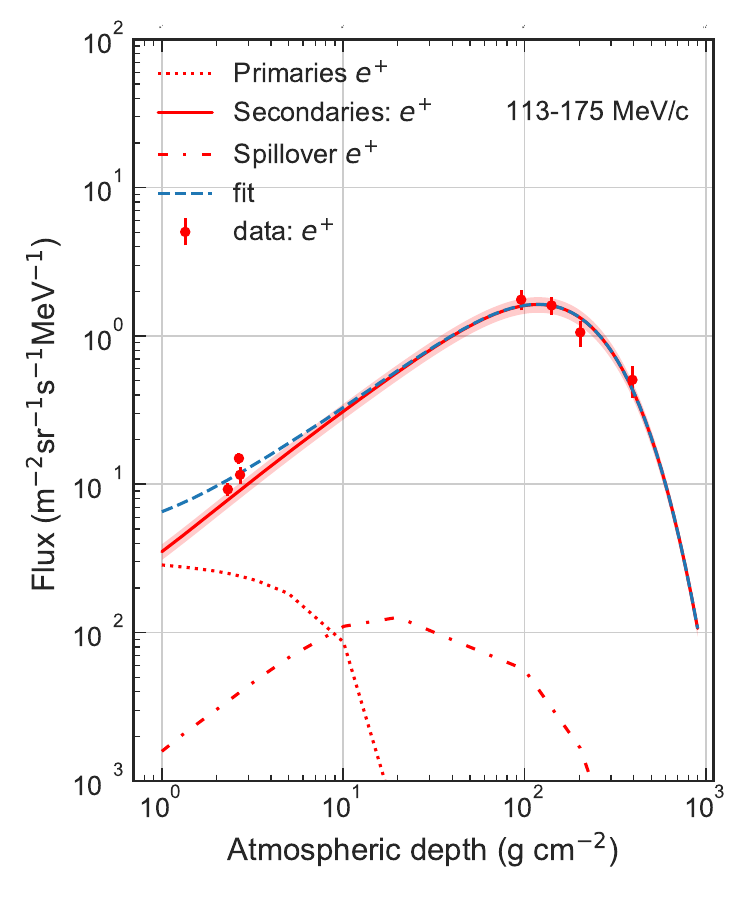}
\includegraphics[page=15,width=0.45\textwidth]{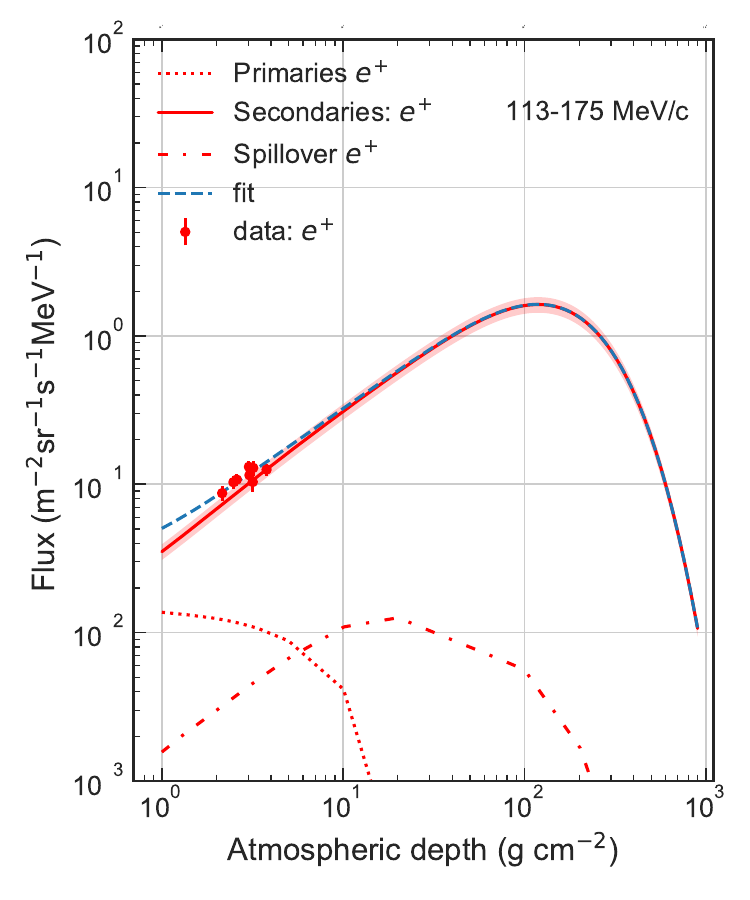}
\caption{(Left) Growth curves fit for positrons in the energy bin 73--113 MeV at daytime. The filled circles represent the flight data, and the dotted, solid and dash-dotted lines contributions from primary, atmospheric secondary and spillover positrons, respectively. (Right) Nighttime growth curves for positrons for the same energy bins.}
\label{fig:GCFIT}
\end{figure}

To extract the flux at ToA at each energy bin, we implement a simple linear least squares fit, slightly modified from~\cite{fulks_solar_1975}, in which we consider the three contributions to the data:
\begin{equation}
\textrm{data} (d) = a \times \textrm{primaries} (d) + b \times \textrm{secondaries} (d)  + \textrm{spillover} (d),
\end{equation}
where $d$ is the atmospheric depth, and $a$ and $b$ the parameters of the fit, and the spillover contribution is iteratively calculated in the analysis. We proceed with an iterative method of fitting the daytime growth curves first. Since the balloon was ascending through the atmosphere during geomagnetic day, and the background dominates at high atmospheric depths (low altitudes), a fit to the entire atmospheric range at daytime is necessary to evaluate the contribution $b$ of the secondaries. For that same energy bin, we then fit the nighttime growth curve, for float altitudes only, fixing the contribution $b$ to the daytime derived value. For energy bins above $\sim$ 300~ MeV, the daytime and nighttime points are combined since they are above the maximum geomagnetic cutoff. The left panel of Fig.~\ref{fig:GCFIT} illustrates this fit method: the three growth curve contributions are fitted to the DT data growth curves from 2 to 900 g.cm$^{-2}$, and the parameters $a$ and $b$ are estimated. The right panel of Fig.~\ref{fig:GCFIT} shows the fit performed for a nighttime bin, for points ranging from 2--4 g.cm$^{-2}$, with parameter $b$ fixed. The fit value of parameter $a$ then corresponds, in  MeV m$^{2}$sr$^{-1}$s$^{-1}$, to the flux at ToA for the given energy bin.

The fits are done in descending order of energy. At the first iteration (647~MeV--1~GeV), the ``\textit{spillover}'' contribution is calculated assuming a specific spectrum above 1 GeV. We initialize the fit to the LEE09 flux \cite{evenson_cosmic_2009} for electrons and positrons combined (all electrons), and scale the flux assuming a positron fraction $\dfrac{e^{+}}{e^{+}+e^{-}}$ = 0.2. Once the flux at ToA is extracted for the first energy bin, the ``\textit{spillover}'' contribution into the lower bins is updated, and the fit routine iterated.

\begin{table}[h!]
\centering\caption {Fit values of parameter $b$}
\begin{tabular}{ccc}
\toprule Mean Energy at ToA (MeV) & \multicolumn{2}{c}{$b$} \\
\cline{2-3}  &   $e^{-}$   & $e^{+}$    \\    \hline
823.23 & 0.991 $\pm$ 0.103  & 0.730 $\pm$ 0.092 \\
532.84 & 0.705 $\pm$ 0.079  & 0.680 $\pm$ 0.090 \\
344.82 & 0.686 $\pm$ 0.079  & 0.781 $\pm$ 0.073 \\
223.09 & 0.766 $\pm$ 0.076  & 0.713 $\pm$ 0.152 \\
144.26 & 0.955 $\pm$ 0.115  & 0.942 $\pm$ 0.114 \\
93.23 & 1.052 $\pm$ 0.134  & 1.033 $\pm$ 0.140 \\
60.18 & 1.045 $\pm$ 0.062  & 1.026 $\pm$ 0.049 \\
38.78 & 1.039 $\pm$ 0.059  & 1.167 $\pm$ 0.046 \\
24.93 & 0.955 $\pm$ 0.080  & 0.997 $\pm$ 0.130 \\
\botrule\end{tabular}
\label{tab:bpar}

\end{table}

This method allows us simultaneously to extract the re-entrant albedo (daytime) and primary cosmic ray spectra flux (nighttime) for electrons and positrons. We present and discuss our results in the next section. 

\section{Results and discussion} \label{sec:results}

\subsection{Systematic uncertainties} \label{subsec:sysuncertainties}

The determination of the extraterrestrial electron and positron fluxes with a balloon-borne instrument at our energy range is complicated by two factors: the residual layer of atmosphere above the payload, and the fact that the balloon was launched during a transition phase in the diurnal geomagnetic cutoff variations. Thus, depending on their energy, ascent points can be of trapped secondary particles.

The main systematic errors arise from uncertainties on the secondary production in the atmosphere. As was shown in Fig.~\ref{fig:GCdata}, MC simulations of protons and alpha particles gave a good agreement with the data at Regener-Pfotzer maximum. However, any deviation in shape of the secondary growth curves can have an important effect on the final spectrum, considering that the primary signal is very close to the background at float altitudes. This is particularly true for positrons in energy bins near 100 MeV, as evident from Fig.~\ref{fig:UnfoldedSpectraTOP}. Three parameters of the secondary production are studied: the choice of H and He LIS, the value of the solar modulation parameter $\phi$, and the scale factor $F_{hn}$ applied to the He spectrum to estimate the contribution of heavier nuclei. Our ``baseline'' spectrum was derived using the LIS parametrized from Voyager data (column 6, Table 3 of~\cite{ghelfi_non-parametric_2016}), assuming $\phi$ = 438 MV and $F_{hn}$=1.445. We first calculate the systematic errors stemming from the choice of LIS, testing the median flux (without Voyager data) from the same reference, as well as the LIS constructed from ~\cite{vos_new_2015}. Given the strong correlation between the choice of an interstellar spectrum and the determination of $\phi$~\citep{herbst_importance_2010}, we must apply a different modulation potential to the LIS taken from~\cite{vos_new_2015}. We take the calibrated value from ~\cite{usoskin_heliospheric_2017}, $\phi_{Uso}$ = 446 MV \footnote{\url{http://cosmicrays.oulu.fi/phi/phi.html}}. For electrons, the systematic uncertainty on the chosen LIS is of the order of 6\% at 25 MeV to 14\% at 145 MeV. For positrons, the effect is equally as important in the energy bins closest to the background of secondaries. This highlights the delicate task of extracting the spectra in the regions of the ``turn-up'', around 100 MeV, at float altitudes.
We then vary the modulation parameter $\phi$ = 438 MV by $\pm$ 50 MV, for our ``baseline" LIS, and find that for electrons the uncertainty is below 5\%, except at 144 MeV (11\%). For positrons it reaches 28\% at 93 MeV.
A 10\% change to the scale factor $F_{hn}$ is studied, taken as a rough estimate of the uncertainty on the parameter: this effect changes the spectra by less than 1\% for both electrons and positrons. 

The effects of the initial hypotheses of the fit are also taken into consideration in the systematic uncertainties: the electron flux as well as the initial value of the positron fraction above 1 GeV were modified using the LEE11 and PAMELA 2009 results, and varying the positron fraction by $\pm$50\%, since the fraction is known to lie between $\sim$ 0.1--0.3~\citep{adriani_time_2015}. The initialization of the fit is found to account for an uncertainty of less than 1\% for both electrons and positrons. 

The systematic uncertainties resulting from the uncertainty of the selection efficiencies $\epsilon_{sel}$ are already included in the unfolding procedure, in addition to any systematic error related to the algorithm's iterations until the convergence criterion is met. As was explained in \textsection \ref{subsec:efficiency}, the inefficiency of the instrument's trigger is also taken into account in the flux calculation. \\

\subsection{Re-entrant albedo spectra below 100 MeV} \label{subsec:resultsToAALb}
\begin{figure}[ht!]
\centering
\includegraphics[width=0.45\textwidth]{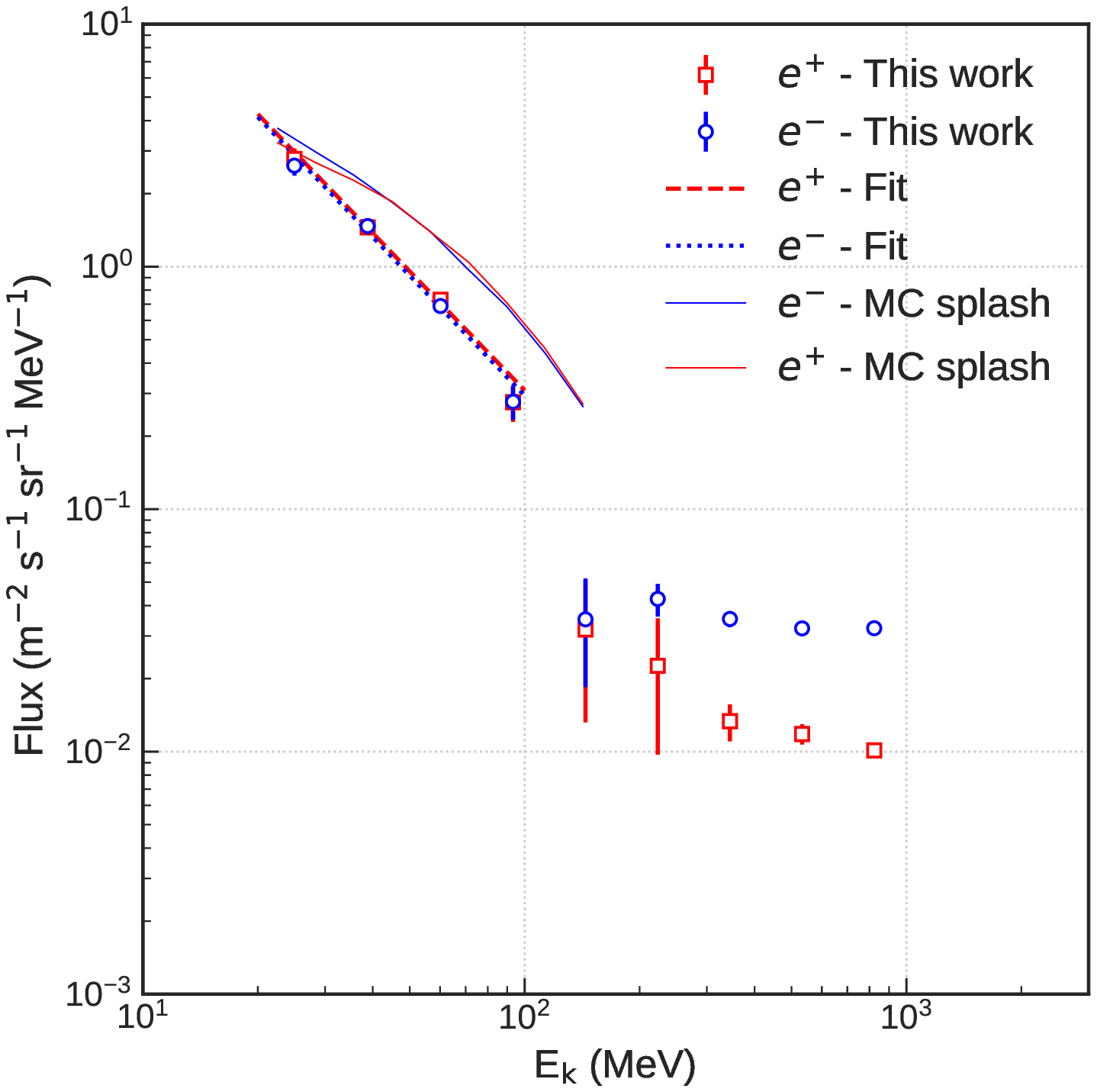}
\includegraphics[width=0.45\textwidth]{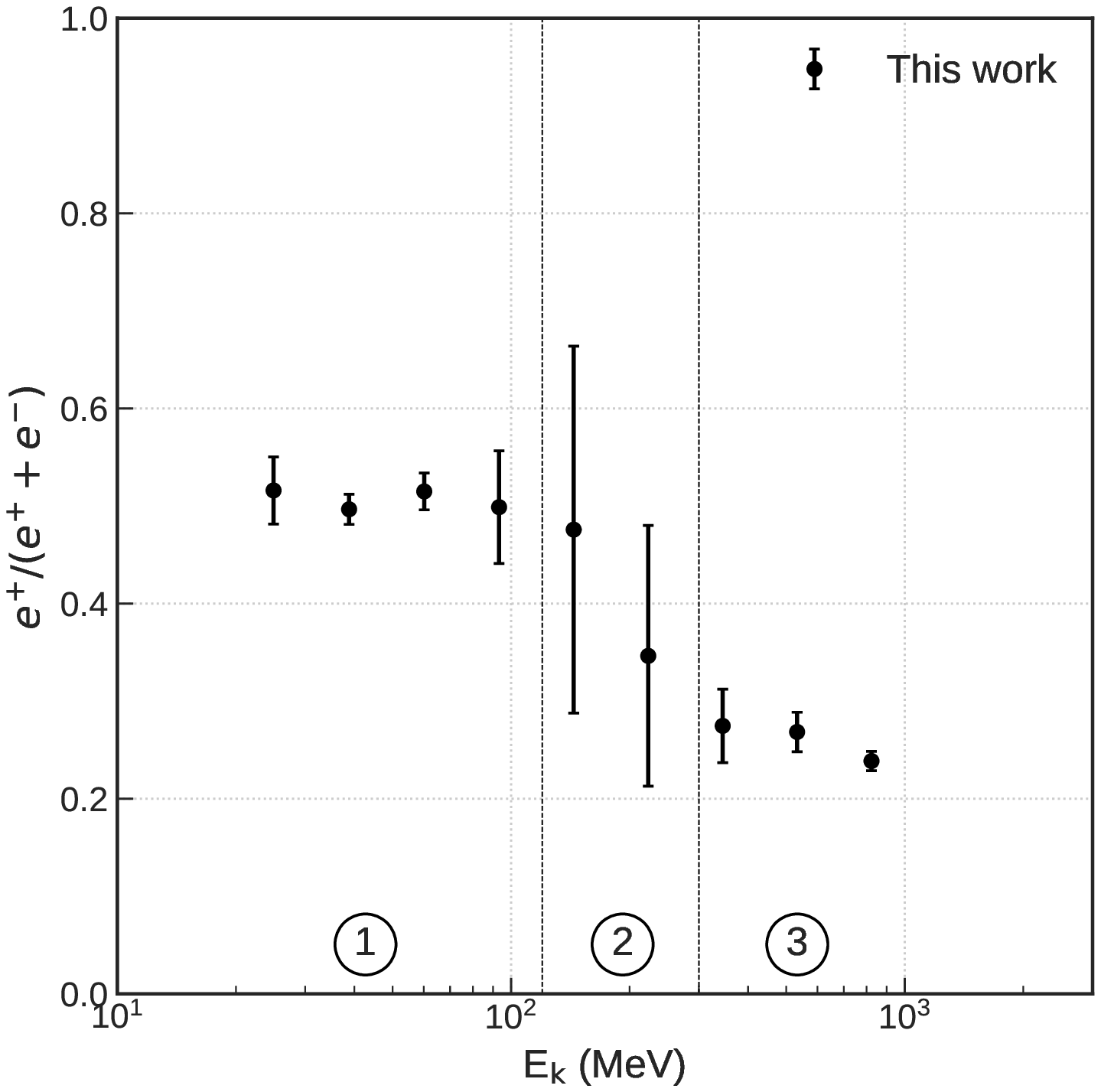}
\caption{(Left) ``Daytime'' spectra of electrons and positrons between 20 MeV and 1 GeV. (Right) ``daytime'' positron fraction at the top of the atmosphere. The energy range 1, below 100 MeV, is dominated by the re-entrant albedo particles. The range 2, between 100 and 300 MeV, is the transition around the geomagnetic cutoff. The range 3 is dominated by primary particles.}
\label{fig:DTspectra}
\end{figure}

The spectra of re-entrant albedo electrons and positrons at ToA are presented in the left panel of Fig.~\ref{fig:DTspectra}. The analysis of the daytime portion of the flight yields a flux of the re-entrant albedo electrons and positrons below 160 MeV (left panel of Fig.~\ref{fig:DTspectra}). This limit comes from the value of the geomagnetic cutoff at the time of the ascent.

Below the 160 MeV cutoff mark (range 1 in the right panel), the measurements in the first hours of flight were primarily of trapped albedo particles. Electrons and positrons above that energy bin, however, were of primary origins (range 3). Range 2 constitutes the transition region of the geomagnetic time, a zone where the origin of the measured particle is somewhat blurrier, in part due to the uncertainties in the geomagnetic simulation performed. These three regions and their spectral implications were already presented in the growth curves of Fig.~\ref{fig:GCdata}.

We fit a simple power-law to the electron and positron spectra below 100 MeV, of the form
\begin{equation}
f(E)=AE^{-\gamma}
\end{equation}
Both fits give a spectral index $\gamma=1.6\pm0.1$, which differs from the results from~\cite{verma_measurement_1967}, who found the re-entrant albedo spectrum to be well fitted with a power-law index $\gamma=1.44\pm0.09$.  We note that that previous measurement had only one data point below 100 MeV, with the highest energy bin extending to 1250 MeV. 

The flux of splash albedo particles from the MC simulation are also visible in the left panel of Fig.~\ref{fig:DTspectra}. As expected from measurements~\citep{verma_measurement_1967}, the spectral index differs from that of the re-entrant component, with $\gamma\sim1.3$.
We observe in the right panel of Fig.~\ref{fig:DTspectra} the clear presence of two regimes of the positron fraction, above and below the cutoff. At higher energies, in range 3, the positron fraction is close to $\sim$ 0.25. Below the cutoff however, the constant positron fraction, very close to 0.5, agrees with the prediction that all re-entrant albedo particles are the secondary products of hadronic interactions of cosmic rays in the atmosphere.

\begin{table}\caption {Flux of re-entrant albedo electron and positron at the top of the atmosphere}
\centering
\begin{tabular}{ccccc}
\toprule
Mean Energy & \phantom{llll}& \multicolumn{3}{c}{Flux at ToA \par (m$^{2}$ sr s MeV)$^{-1}$} \\
\cline{2-5} at ToA (MeV) & \phantom{llll}&   $e^{-}$   & \phantom{llll}& $e^{+}$    \\    \hline
24.93 & & \num{2.61e+00} $\pm$ \num{2.37e-01} & & \num{2.78e+00} $\pm$ \num{2.89e-01}   \\
38.78 & & \num{1.47e+00} $\pm$ \num{7.59e-02} & & \num{1.45e+00} $\pm$ \num{4.96e-02}   \\
60.18 & & \num{6.88e-01} $\pm$ \num{4.13e-02} & & \num{7.30e-01} $\pm$ \num{3.34e-02}   \\
93.23 & & \num{2.77e-01} $\pm$ \num{4.33e-02} & & \num{2.76e-01} $\pm$ \num{4.71e-02}   \\
\botrule
\end{tabular}
\end{table}

\subsection{Electron and positron spectra} \label{subsec:resultsToA}

\begin{figure}
\centering
\includegraphics[width=0.45\textwidth]{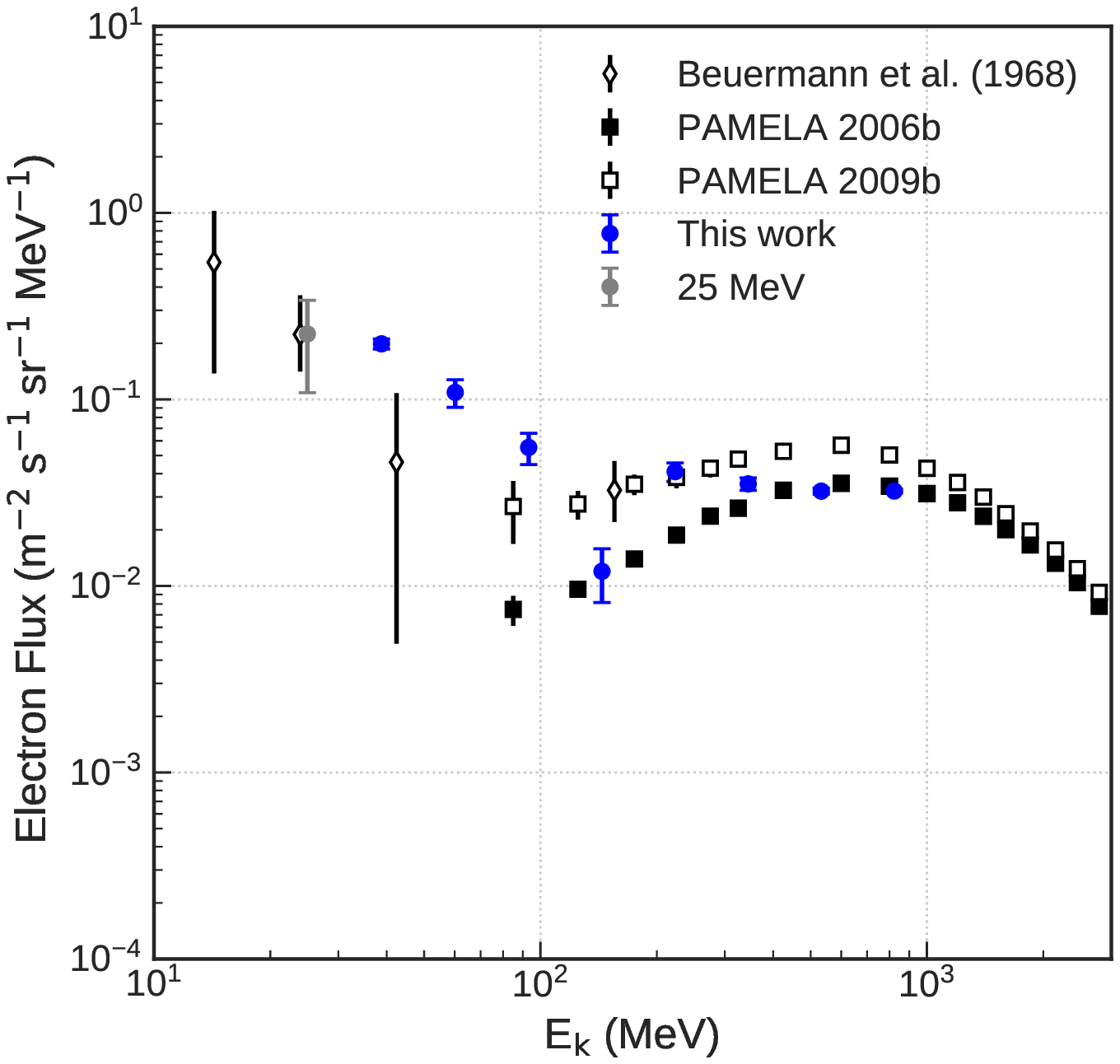}
\includegraphics[width=0.45\textwidth]{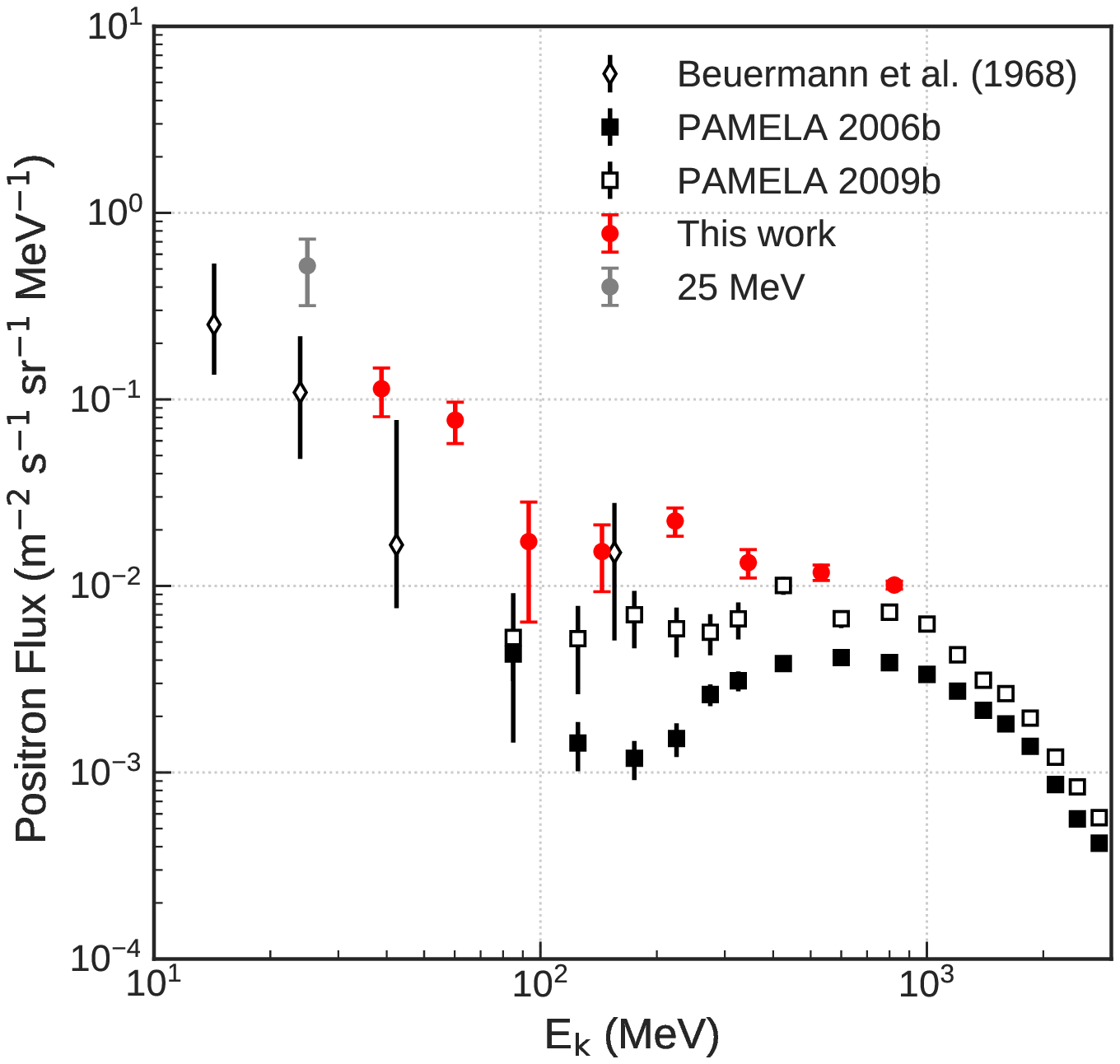}
\caption{(Left) Primary spectrum of cosmic ray electrons between 20 MeV and 1 GeV. (Right) Primary spectrum of cosmic ray positrons in the same energy range. The index "b" corresponds to measurements made during the second semester of the year. PAMELA electron and positron data are taken from~\cite{adriani_time_2015}, and~\cite{aslam_modeling_2019}, respectively.}
\label{fig:NTspectra}
\end{figure}

The primary electron and positron spectra measured by the AESOP-Lite instrument are shown in the two panels of Fig.~\ref{fig:NTspectra}, from 30 MeV to 1 GeV. The points at the lowest energy bin are shown in gray because of inconsistencies found when unfolding the spectra using two different reconstruction algorithms: this reflects the difficulty of extracting the flux at ToA during nighttime, while normalizing the secondary growth curve to the daytime ascent. The low statistics of the ascent phase in the 20--30 MeV edge bin cause greater uncertainties in the unfolding procedure and the growth curve fit.

Both of the electron and positron spectra display a ``turn-up'', the name we give to the transition region around 80--100 MeV where the spectral index changes and becomes negative at lower energies: this had previously been observed in the all electron spectrum measured by the LEE payload~\citep{fulks_solar_1975,evenson_quantitative_1983,evenson_cosmic_2009}, and had been hinted at in PAMELA data down to 80 MeV~\citep{adriani_time_2016,aslam_modeling_2019}. This behavior is also revealed in our positron spectrum, despite the uncertainties in the data points, as explained above.

The comparison between AESOP-Lite's and PAMELA's electron and positron measurements is qualitatively consistent with the knowledge we have of charge-sign dependent solar modulation. In the second semester of the year 2009 (``2009b"), the solar modulation parameter $\phi$ as measured by the methodology of~\cite{ghelfi_neutron_2017} was $\phi$ $\sim$ 439~MV, while it was $\phi$ $\sim$ 539~MV in 2006b. The lower solar modulation environs of 2009 help explain the higher amplitude in the electron and positron fluxes that PAMELA measured in 2009 compared to 2006, during the same A- solar polarity cycle, taking into consideration the well-known anti-correlation between the solar activity and the cosmic ray flux at 1 AU. Our 2018 flight took place during a very low solar minimum, with $\phi$ $\sim$ 438~MV, a value similar to that present during PAMELA's data taking. However the solar epoch had changed from an A- polarity in 2009 to an A+ one in 2018. This has a notable impact on the propagation of charged particles: in a positive cycle, positrons reach the Earth with greater ease than electrons, having traveled via the polar regions of the heliosphere, whereas electrons encounter more of the gradient and curvature of the wavy HCS present in the helio-equatorial regions they traversed~\citep{jokipii_effects_1977}, their flux thus more suppressed. We notice that at energies above 100~MeV, the electron flux measured by AESOP-Lite is lower than that reported by PAMELA in 2009, which can likely be explained by the polarity reversal. Conversely, the positron flux recorded by our instrument is higher in amplitude than PAMELA's, consistent with the fact that the A+ epoch favors positively charged particles more so than the A- epoch does.

At lower energies, the only comparable separate  measurements of electrons and positrons are from~\cite{beuermann_cosmic-ray_1969}, in which a balloon-borne magnetic spectrometer measured particles down to 12 MeV; their data suggests a similar power-law form.

\begin{figure}[ht!]
\centering
\includegraphics[width=0.5\textwidth]{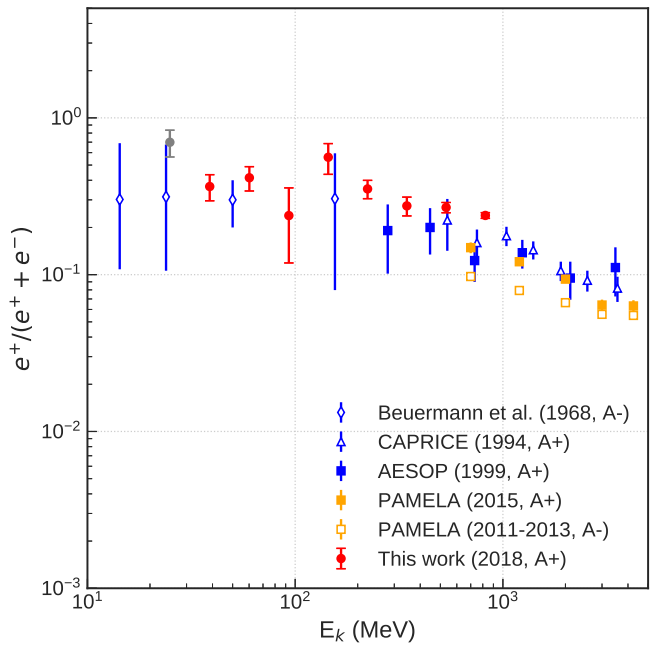}
\caption{Positron fraction of primary cosmic rays.}
\label{fig:PosFraction}
\end{figure}

The positron fraction of the primary cosmic ray spectrum is presented in Fig.~\ref{fig:PosFraction}. 
Above 200 MeV, the fraction suggests a rise with decreasing energy, a trend previously displayed in results from the PAMELA, AESOP and CAPRICE94 experiments~\citep{adriani_time_2016,clem_solar_1996,boezio_cosmic-ray_2000}, to name a few. The solar polarity cycle appears to have an effect on the positron fraction: for instance, the measurement by AESOP-Lite at 1 GeV in a A+ epoch is significantly higher than the one made by PAMELA in A-. This temporal variation caused by the charge-sign dependent solar modulation had previously been observed by PAMELA and AMS-02~\citep{adriani_time_2016,aguilar_observation_2018}, and reproduced at higher energy using numerical transport codes~\citep{potgieter_charge-sign_2014}. We note that the fraction we measured at higher energy is also significantly greater than the one observed by PAMELA in a similar polarity cycle, presumably because the Sun's activity was at a minimum in 2018.

At first glance, the positron fraction appears to be flat from 30 MeV to 200 MeV, plateauing at $\sim$ 0.3, indicating that the flux consists of a mixture of ``primary'' GCR electrons, and ``secondary'' GCR positrons produced in the propagation of cosmic rays~\citep{moskalenko_production_1998}. An interesting observation, relevant to the study of the charge-sign modulation and the effects of drift at low energy, is the apparent agreement between our data points and those collected by~\cite{beuermann_cosmic-ray_1969}. Those measurements were made in 1968, during an A- polarity cycle, whereas ours were made during an A+ cycle. This suggests that diffusion might dominate over drift effects at lower energies, whereas a different mechanism is at play above 200 MeV.

\begin{figure}
\centering
\includegraphics[width=0.45\textwidth]{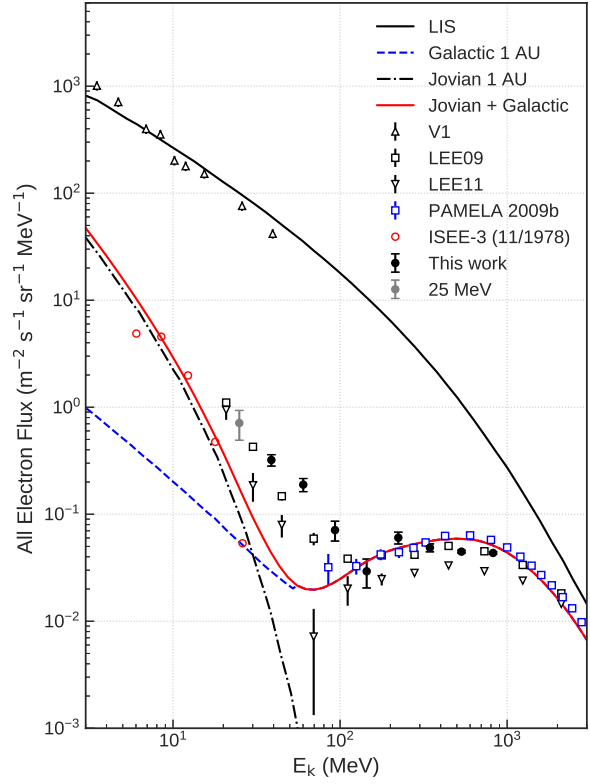}
\caption{Primary spectrum of cosmic ray the all electrons between 20 MeV and 1 GeV, shown with Voyager all electrons outside the heliosphere.}
\label{fig:AllElecspectra}
\end{figure}

Fig.~\ref{fig:AllElecspectra} shows the all-electron spectrum, alongside the two last measurements completed by LEE09 and LEE11. Below 50 MeV, the Jovian magnetosphere also becomes an important source of electrons~\citep{vogt_jovian_2018}, as detected by the ISEE-3 satellite mission at 1~AU (red circles in the figure)~\citep{moses_jovian_1987}. Elaborate 3D numerical transport codes have been developed over the past decades~\citep{potgieter_modulation_2015,vos_new_2015,aslam_modeling_2019,bisschoff_new_2019}, in which the different processes of the theory of solar modulation are included: namely, the convection, adiabatic deceleration, drift and diffusion of charged particles in the solar wind. The dashed blue lines represents the model of~\cite{potgieter_modulation_2015}, showing the LIS Voyager 1 spectra (solid black line) propagated through the heliosphere. PAMELA electron observations were used to tune model parameters. In dashed black is the prediction of the modulated Jovian spectrum, whereas the red solid line projects the expected electron spectrum at Earth for a given solar epoch and modulation potential~\citep{nndanganeni_global_2018}. The crossover between the GCR electrons and the Jovian electrons is estimated to happen at 30 MeV, according to~\cite{nndanganeni_global_2018}. The final prediction of the all electron flux at 1 AU notably involves a ``turn-up'' around 80 MeV, and a negative power-law behavior below. From the combined study of AESOP-Lite and LEE data, the energy at which the minimum occurs is shifting, as LEE11 results coincidence with the model, whereas AESOP-Lite and LEE09 results do not. 

In fact, a dedicated study of the solar modulation of electrons and positrons is needed in order to characterize the interplay of Galactic and Jovian of electron sources as well as drift and diffusion effects below 100 MeV. Diffusion and drift coefficients are proportional to the mean free paths (MFP). For electrons, the parallel and perpendicular MFPs, which govern the diffusive process, are assumed to be rigidity-independent below a yet vague threshold ($\sim$ 100 MeV) ~\cite{bieber_proton_1994,droge_solar_2003,potgieter_modulation_2015}. The addition of our data set can offer a glimpse into the behavior of electron and positron cosmic rays in a poorly observed energy regime. The contemporary measurements of Voyager 1 and 2, progress in the numerical modeling, and the planned future missions of the AESOP-Lite instrument create a unique opportunity to finally resolve the origin of the low-energy electron and positron spectra on Earth.

\begin{table}\caption {Primary electron and positron flux at the top of the atmosphere}
\centering
\begin{tabular}{ccccc}
\toprule
Mean Energy  & \phantom{llll} & \multicolumn{3}{c}{Flux at ToA \par (m$^{2}$ sr s MeV)$^{-1}$} \\
\cline{3-5} at ToA (MeV)& \phantom{llll}  &   $e^{-}$   & \phantom{llll} & $e^{+}$      \\    \hline
24.93 & & \num{2.24e-01} $\pm$ \num{1.16e-01} & & \num{5.21e-01} $\pm$ \num{2.02e-01}   \\
38.78 & & \num{1.99e-01} $\pm$ \num{1.22e-02} & & \num{1.14e-01} $\pm$ \num{3.33e-02}   \\
60.18 & & \num{1.09e-01} $\pm$ \num{1.84e-02} & & \num{7.74e-02} $\pm$ \num{1.93e-02}   \\
93.23 & & \num{5.53e-02} $\pm$ \num{1.05e-02} & & \num{1.73e-02} $\pm$ \num{1.09e-02}   \\
144.26 & & \num{1.20e-02} $\pm$ \num{3.83e-03} & & \num{1.53e-02} $\pm$ \num{5.98e-03}  \\
223.09 & & \num{4.10e-02} $\pm$ \num{4.69e-03} & & \num{2.23e-02} $\pm$ \num{3.84e-03}   \\
344.82 & & \num{3.52e-02} $\pm$ \num{2.67e-03} & & \num{1.33e-02} $\pm$ \num{2.31e-03}   \\
532.84 & & \num{3.22e-02} $\pm$ \num{1.23e-03} & & \num{1.18e-02} $\pm$ \num{1.13e-03}   \\
823.23 & & \num{3.23e-02} $\pm$ \num{7.78e-04} & & \num{1.01e-02} $\pm$ \num{4.97e-04}   \\
\botrule
\end{tabular}
\end{table}

\begin{table}
\centering
\caption {Positron fraction}
\begin{tabular}{cc}
\toprule\hline
Mean Energy at ToA (MeV) & $\dfrac{e^{+}}{e^{+}+e^{-}}$ \\ 
\hline
24.93 & \num{6.99e-01} $\pm$ \num{1.36e-01}  \\
38.78 & \num{3.65e-01} $\pm$ \num{6.91e-02}  \\
60.18 & \num{4.15e-01} $\pm$ \num{7.33e-02}  \\
93.23 & \num{2.38e-01} $\pm$ \num{1.19e-01}  \\
144.26 & \num{5.61e-01} $\pm$ \num{1.24e-01}  \\
223.09 & \num{3.52e-01} $\pm$ \num{4.72e-02}  \\
344.82 & \num{2.75e-01} $\pm$ \num{3.77e-02}  \\
532.84 & \num{2.68e-01} $\pm$ \num{2.02e-02}  \\
823.23 & \num{2.39e-01} $\pm$ \num{9.95e-03}  \\
\botrule
\end{tabular}
\end{table}

\acknowledgments

The authors would like to thank Matthew Collins, Forest Martinez-McKinney, Serguei Kachiguine, and Yang Zhou for their help in the design, construction and integration of the instrument.
We thank Chris Field and CSBF for their support during the integration period in Palestine, Texas, and for the successful balloon flight. 
We thank Esrange for their support during the flight campaign. 
This work is supported by NASA awards NNX15AL32G and 80NSSC19K0746, and the Bartol Research Institute.

\clearpage

\appendix

\section{Full electron spectra of LEE instrument from 2009 and 2011 flights}

Two balloon flights carrying the LEE instrument took place during the dates May 16-21, 2009 and May 26-31, 2011 from the Esrange Space Center near Kiruna, Sweden.
Upon reaching the stratosphere the balloons rode the summer Arctic polar vortex across the Atlantic Ocean, Greenland and the Baffin Bay into the northern regions of Canada where the flights were terminated.
While both flights occurred during A- solar minimum, the 2009 flight took place during
exceptionally low solar modulation level, the lowest during the history of the neutron monitors.
The long duration exposure combined with low modulation level allowed the complete electron spectrum from 20 MeV to 5 GeV to be observed for the very first time. Flight data were analyzed using the same method outlined in~\cite{fulks_solar_1975}.

\begin{table}[ht!]
\centering
\caption {Full electron spectra of LEE instrument from 2009 and 2011 flights}
\begin{tabular}{ccccc}
\toprule
Mean Energy & \phantom{llll}  & \multicolumn{3}{c}{Flux $e^{+}+e^{-}$ at ToA \par (m$^{2}$ sr s MeV)$^{-1}$} \\
\cline{3-5} at ToA (MeV)& \phantom{llll} &   May 16-21, 2009  & \phantom{llll}  & May 26-31, 2011      \\    \hline
\num{2.080E+01} & & \num{1.105E+00} $\pm$ \num{7.895E-02} & &\num{9.421E-01} $\pm$ \num{1.796E-01}\\
\num{3.010E+01} & & \num{4.266E-01} $\pm$ \num{2.303E-02} & &\num{1.868E-01} $\pm$ \num{5.602E-02}\\
\num{4.480E+01} & & \num{1.474E-01} $\pm$ \num{1.315E-02} & &\num{7.949E-02} $\pm$ \num{1.893E-02}\\
\num{6.950E+01} & & \num{5.909E-02} $\pm$ \num{7.851E-03} & &\num{7.185E-03} $\pm$ \num{5.860E-03}\\
\num{1.110E+02} & & \num{3.841E-02} $\pm$ \num{1.606E-03} & &\num{2.016E-02} $\pm$ \num{6.208E-03}\\
\num{1.775E+02} & & \num{4.125E-02} $\pm$ \num{1.610E-03} & &\num{2.475E-02} $\pm$ \num{3.174E-03}\\
\num{2.790E+02} & & \num{4.180E-02} $\pm$ \num{1.533E-03} & &\num{2.828E-02} $\pm$ \num{1.701E-03}\\
\num{4.460E+02} & & \num{5.055E-02} $\pm$ \num{1.056E-03} & &\num{3.316E-02} $\pm$ \num{6.451E-04}\\
\num{7.310E+02} & & \num{4.506E-02} $\pm$ \num{6.700E-04} & &\num{2.933E-02} $\pm$ \num{4.318E-04}\\
\num{1.240E+03} & & \num{3.356E-02} $\pm$ \num{3.011E-04} & &\num{2.394E-02} $\pm$ \num{1.531E-04}\\
\num{2.110E+03} & & \num{1.821E-02} $\pm$ \num{2.069E-04} & &\num{1.452E-02} $\pm$ \num{1.147E-04}\\
\num{3.475E+03} & & \num{7.741E-03} $\pm$ \num{6.988E-05} & &\num{6.886E-03} $\pm$ \num{1.135E-04}\\
\num{5.800E+03} & & \num{2.671E-03} $\pm$ \num{5.344E-05} & &\num{2.398E-03} $\pm$ \num{3.122E-05}\\
\num{1.005E+04} & & \num{6.616E-04} $\pm$ \num{2.159E-05} & &\num{6.128E-04} $\pm$ \num{1.735E-05}\\
\num{1.865E+04} & & \num{9.470E-05} $\pm$ \num{3.509E-05} & &\num{8.599E-05} $\pm$ \num{5.721E-06}\\
\botrule
\end{tabular}
\end{table}

\bibliography{ApJ_AESOPLITE}
\bibliographystyle{aasjournal}

\end{document}